\documentclass[prd,reprint,aps,preprintnumbers,amsmath,amssymb,showpacs,superscriptaddress,
nofootinbib]{revtex4-1}
\usepackage[hyperindex,breaklinks]{hyperref}
\usepackage{graphicx,bm,epstopdf}

\def\lsim{\mathrel{\rlap{\lower4pt\hbox{\hskip1pt$\sim$}}
    \raise1pt\hbox{$<$}}}                % less than or approx. symbol
\def\gsim{\mathrel{\rlap{\lower4pt\hbox{\hskip1pt$\sim$}}
    \raise1pt\hbox{$>$}}}                % greater than or approx. symbol
\def\MeV{{~\rm MeV}}
\def\GeV{{~\rm GeV}}
\def\TeV{{~\rm TeV}}

\begin{document}
\preprint{SCIPP-12/04}

\title{Diphotons from Tetraphotons in the Decay of a 125 GeV Higgs at the LHC}

\author{Patrick Draper}
\email{pidraper@ucsc.edu}
\affiliation{Santa Cruz Inst. for Particle Physics, Univ. of California, 
Santa Cruz, CA 95064, USA}
\author{David McKeen}
\email{mckeen@uvic.ca}
\affiliation{Department of Physics and Astronomy, University of Victoria, 
Victoria, BC V8P 5C2, Canada}

%%%%%%%%%%%%%%%%%%%%%%%%%%%%%%%%%%%%%%%%%%%%%%%%%%%%%%%%%%%%%%%%%%%%%%%%%%%%%%%%
%%%%%%%%%%%%%%%%%%%%%%%%%%%%%%%%%%%%%%%%%%%%%%%%%%%%%%%%%%%%%%%%%%%%%%%%%%%%%%%%
%%%%%%%%%%%%%%%%%%%%%%%%%%%%%%%%%%%%%%%%%%%%%%%%%%%%%%%%%%%%%%%%%%%%%%%%%%%%%%%%

\begin{abstract} 
Recently the ATLAS and CMS experiments have presented data hinting at the presence of a Higgs boson at $m_h\simeq125$ GeV. The best-fit $h\rightarrow\gamma\gamma$ rate averaged over the two experiments is approximately $2.1\pm0.5$ times the Standard Model prediction. We study the possibility that the excess relative to the Standard Model is due to $h\rightarrow aa$ decays, where $a$ is a light pseudoscalar that decays predominantly into $\gamma\gamma$. Although this process yields $4\gamma$ final states, if the pseudoscalar has a mass of the order tens of MeV, the two photons from each $a$ decay can be so highly collimated that they may be identified  as a single photon. Some fraction of the events then contribute to an effective $h\rightarrow\gamma\gamma$ signal. We study the constraints on the parameter space where the net $h\rightarrow\gamma\gamma$ rate is enhanced over the Standard Model by this mechanism and describe some simple models that give rise to the pseudoscalar-photon interaction.  Further tests and prospects for searches in the near future are discussed.
\end{abstract}  
\maketitle  

%%%%%%%%%%%%%%%%%%%%%%%%%%%%%%%%%%%%%%%%%%%%%%%%%%%%%%%%%%%%%%%%%%%%%%%%%%%%%%%%
%%%%%%%%%%%%%%%%%%%%%%%%%%%%%%%%%%%%%%%%%%%%%%%%%%%%%%%%%%%%%%%%%%%%%%%%%%%%%%%%
%%%%%%%%%%%%%%%%%%%%%%%%%%%%%%%%%%%%%%%%%%%%%%%%%%%%%%%%%%%%%%%%%%%%%%%%%%%%%%%%

\begin{section}{Introduction}
Determining the nature of electroweak symmetry breaking (EWSB) has been a primary goal of particle physics for several decades.  In the Standard Model (SM), EWSB occurs when the neutral component of a single scalar weak isospin doublet possesses a vacuum expectation value (VEV).  The Higgs boson corresponds to the physical excitations of this field.  Recent experimental advances make it clear that we are entering a new phase in searches for the Higgs boson, with tantalizing excesses appearing in several SM-like Higgs search channels at both the CERN Large Hadron Collider (LHC) and the Fermilab Tevatron.  These hints point to a relatively light state with a mass in the range $122.5\lesssim m_h\lesssim 127.5$~GeV~\cite{MoriondATLAS, MoriondCMS}.  In the SM, such a light Higgs boson has an extremely narrow width of the order $10^{-5}\times m_h$.  The tiny width of the SM Higgs makes it a sensitive probe of new physics (NP) beyond the SM, especially in the only loosely constrained scalar sector~\cite{Chang:2008cw}.  New states coupled to a SM-like Higgs can have appreciable effect on its decays without ruining the excellent agreement of the SM with data observed elsewhere thus far (see, e.g.,~\cite{Dermisek:2007yt}).

In fact, the excesses seen at the LHC and Tevatron may already be pointing toward interesting deviations from the SM predictions for the Higgs branching ratios. The largest statistical power thus far comes from the searches for $h\rightarrow\gamma\gamma$ at the LHC, where the ATLAS experiment finds a 2.8$\sigma$ excess at $m_h\simeq 126$ GeV with a best-fit signal strength relative to the SM of approximately $2.0\pm0.8$~\cite{ATLAS:2012ad,MoriondATLAS}. The CMS experiment finds a 3.1$\sigma$ excess in $h\rightarrow\gamma\gamma$ at $m_h\simeq 124$ GeV with a best-fit signal strength relative to the SM of approximately $2.1\pm0.6$~\cite{Chatrchyan:2012tw,MoriondCMS}. Under the assumption that these excesses  are due to the same new particle, we can naively combine the diphoton rates from the two experiments and estimate 
\begin{align}
\mbox{Rate}(h\rightarrow\gamma\gamma)/\mbox{SM} \simeq 2.1\pm0.5.
\end{align}
This estimate is not rigorous and the error bar is far from conclusive. However, the uncertainties will continue to decrease as data accumulate, and the central value may very well remain high. Therefore, it is of interest to classify models that can alter the $h\rightarrow\gamma\gamma$ signal and study their other predictions. A number of recent papers have contributed to this program, including discussions of the effects of superpartners on the diphoton rate~\cite{Carena:2011aa,*Heinemeyer:2011aa,*Arbey:2011un,*Cao:2012fz,*Vasquez:2012hn}, the effects of more general new fermion and scalar states~\cite{Barger:2012hv}, the effects of singlet-doublet mixing~\cite{Ellwanger:2011aa,*Kang:2012tn}, the predictions in the case of minimal universal extra dimensions~\cite{Belanger:2012zg}, interference effects from charged Higgs contributions~\cite{Arhrib:2011vc,*Arhrib:2012ia}, and the possibility that the signal is due to the Randall-Sundrum radion~\cite{Cheung:2011nv}, to name a few.

In this paper we consider a different scenario in which a 125 GeV Higgs boson can appear to have a larger branching ratio into $\gamma\gamma$. We introduce a very light pseudoscalar of mass in the range $m_a\in(\sim10\MeV$, $\sim m_\pi)$, and we allow the Higgs boson to decay in the channel $h\rightarrow aa$. The pseudoscalars produced in these decays are extremely boosted in the lab frame, causing their decay products to be quite collimated.  Furthermore, for such light pseudoscalars, there are no kinematically available hadronic decay modes. As first studied in~\cite{Dobrescu:2000jt}, the decay $a\rightarrow\gamma\gamma$ induced by the effective coupling $aF_{\mu\nu}\tilde{F}^{\mu\nu}$ can easily dominate, leading to a $h\rightarrow aa\rightarrow4\gamma$ signature (in contrast to the ``buried Higgs" scenario where the dominant decays are hadronic~\cite{Berger:2002vs,*Bellazzini:2009xt,*Falkowski:2010hi}).  Because of the large boost for the pseudoscalars produced in $h$ decays, the photon pairs from each $a$ can be so highly collimated that a significant fraction may be identified as single photons, even in the ATLAS and CMS detectors which are very good at distinguishing two closely separated photons (e.g. originating from $\pi^0$ or $\eta$ decays).

A related scenario, although not in the context of Higgs boson decays, was considered in~\cite{Toro:2012sv}, which studied decays of a new heavy vector through pseudoscalars to extremely collimated ``photon jets." Since their vector mass is much larger than 125~GeV, they consider much heavier pseudoscalars. Our analysis of the conditions necessary for photon jets to be identified as single photons is complementary to that of~\cite{Toro:2012sv} and provides further motivation for the detailed study of photon jets. Previously, the $h\rightarrow aa\rightarrow 4\gamma$ scenario was considered in~\cite{Dobrescu:2000jt,Dermisek:2007yt,Chang:2006bw} and, in particular, the relevance of this channel to $h\rightarrow\gamma\gamma$ searches at the Tevatron was studied in~\cite{Dobrescu:2000jt}. While Ref.~\cite{Dobrescu:2000jt} focused mainly on pseudoscalars heavier than 200 MeV and heavier Higgs bosons, we will find that a significant $4\gamma\rightarrow2\gamma$ fake rate at the LHC with a 125 GeV Higgs boson requires a pseudoscalar lighter than about the mass of the $\pi^0$. In contrast to previous works, we also study the interplay between the net expected $h\rightarrow\gamma\gamma$ signal and the signal in other Higgs search channels, provide an analysis of the compatibility of the light pseudoscalar scenario with the current Higgs data from the LHC and Tevatron, and study in detail the constraints from LEP and low-energy experiments.  For further discussion of Higgs decays to light pseudoscalars, see, e.g.,~\cite{Dobrescu:2000yn}.

This paper is organized as follows. In Sec.~\ref{sec:model} we discuss the basic features of the model. In Sec.~\ref{sec:signal} we analyze the changes in the expected rates for SM Higgs search channels at the LHC. We give a detailed estimate of the probability for a collimated photon pair to be identified as a single photon, and perform a basic statistical analysis of the constraints on the light pseudoscalar parameter space coming from current Higgs searches at the Tevatron and LHC. In Sec.~\ref{sec:directconstr} we study the constraints following from the muon anomalous magnetic moment, direct searches at LEP for $e^+e^-\rightarrow\gamma+{\rm inv.}$, meson decays, beam dump experiments, and nuclear physics. In Sec.~\ref{sec:models} we survey model building for the light $a$ and its coupling to photons, and consider the possibility that the latter is generated by $\tau$ loops, or by $a-\pi^0$ mixing. In Sec.~\ref{sec:outlook} we discuss the prospects for directly producing pseudoscalars in the mass and coupling range of interest at Primakoff-type experiments.  In Sec.~\ref{sec:conclusions} we conclude, and an Appendix generalizes our main results to the case where $a$ has a substantial invisible branching fraction.

\end{section}

%%%%%%%%%%%%%%%%%%%%%%%%%%%%%%%%%%%%%%%%%%%%%%%%%%%%%%%%%%%%%%%%%%%%%%%%%%%%%%%%
%%%%%%%%%%%%%%%%%%%%%%%%%%%%%%%%%%%%%%%%%%%%%%%%%%%%%%%%%%%%%%%%%%%%%%%%%%%%%%%%
%%%%%%%%%%%%%%%%%%%%%%%%%%%%%%%%%%%%%%%%%%%%%%%%%%%%%%%%%%%%%%%%%%%%%%%%%%%%%%%%

\begin{section}{Model}
\label{sec:model}
In addition to the SM, we consider a real pseudoscalar $a$ which is the pseudo Nambu-Goldstone boson of some spontaneously broken approximate global symmetry.\footnote{A frequently-discussed model containing a light PNGB is the NMSSM in the approximate PQ- or R-symmetric limits. In that model the pseudoscalar has sufficiently large fermionic couplings that in the mass range we consider, it is completely ruled out by low-energy experiments such as those considered in Sec.~\ref{sec:directconstr}~\cite{Andreas:2010ms}.} It has a small mass $m_a$ coming from small explicit breaking of the symmetry, and it interacts with the SM via
\begin{align}
{\cal L}_{\rm int}&=\frac{1}{\Lambda^2}\left(\partial^\mu a\right)^2H^\dagger H-\frac{e^2}{4M}a\,F^{\mu\nu}\tilde F_{\mu\nu},
\label{eq:L_int}
\end{align}
where $H$ is the SM Higgs doublet, $F^{\mu\nu}$ is the photon field strength, and $e$ is the positron charge. $\Lambda$ and $M$ are scales describing the strength of the higher dimensional operators. 

Expanding around the Higgs VEV $v\simeq 246\GeV$, $H^0=\left(v+h\right)/\sqrt{2}$, leads to a rate for the Higgs to decay to pseudoscalars of
\begin{align}
\Gamma\left(h\to aa\right)&=\frac{v^2m_h^3}{32\pi \Lambda^4}
\\
&=1.18\MeV \left(\frac{m_h}{125~{\rm GeV}}\right)^3\left(\frac{\Lambda}{\TeV}\right)^{-4},
\nonumber
\end{align}
where we have assumed that $m_a\ll m_h$.  Using Eq.~(\ref{eq:L_int}) we can also calculate the rate for $a$ to decay to $\gamma\gamma$,
\begin{align}
&\Gamma\left(a\to\gamma\gamma\right)=\frac{\pi\alpha^2m_a^3}{4 M^2}
\\
&~~~~=2.68\times10^{-8}\MeV\left(\frac{M}{10\GeV}\right)^{-2}\left(\frac{m_a}{40\MeV}\right)^3.
\nonumber
\end{align}
Pseudoscalars produced by the decay of $125\GeV$ Higgs at rest then have a decay length of
\begin{align}
\gamma c\tau&\simeq1.15~{\rm cm}~\left(\frac{M}{10\GeV}\right)^2\left(\frac{m_a}{40\MeV}\right)^{-4}
\label{eq:declength}
\\
&~~~~~~~~~~~~~~~~~~~~~~~~~~~~~~~~\times\left(\frac{m_h}{125~{\rm GeV}}\right).
\nonumber
\end{align}
If we require $\gtrsim90\%$ of the decays to occur inside the electromagnetic calorimeters of ATLAS and CMS ($\simeq 1$m radius), $\gamma c\tau$ should be less than about a half meter.  We can express the scale $M$ in terms of the decay length, $m_a$, and $m_h$,
\begin{align}
M&=9.3~{\rm GeV}~\left(\frac{\gamma c\tau}{1~{\rm cm}}\right)^{1/2}\left(\frac{m_a}{40\MeV}\right)^{2}
\label{eq:scale_M}
\\
&~~~~~~~~~~~~~~~~~~~~~~~~~~~~~~~~\times\left(\frac{m_h}{125\GeV}\right)^{-1/2}.
\nonumber
\end{align}

In the Appendix, we expand this simple model to allow for $a$ to additionally decay invisibly.  We discuss the benefits and further constraints that this entails there.
\end{section}

%%%%%%%%%%%%%%%%%%%%%%%%%%%%%%%%%%%%%%%%%%%%%%%%%%%%%%%%%%%%%%%%%%%%%%%%%%%%%%%%
%%%%%%%%%%%%%%%%%%%%%%%%%%%%%%%%%%%%%%%%%%%%%%%%%%%%%%%%%%%%%%%%%%%%%%%%%%%%%%%%
%%%%%%%%%%%%%%%%%%%%%%%%%%%%%%%%%%%%%%%%%%%%%%%%%%%%%%%%%%%%%%%%%%%%%%%%%%%%%%%%

\begin{section}{Higgs Signal}
\label{sec:signal}
The interactions described in Sec.~\ref{sec:model} do not appreciably affect the production cross section of the SM Higgs boson at the LHC. In this section we study the effects of the decays $h\rightarrow aa$, $a\rightarrow\gamma\gamma$ on the expected Higgs-to-diphotons rate. We also estimate the constraints on such decays from existing SM Higgs searches at the Tevatron and LHC.

%%%%%%%%%%%%%%%%%%%%%%%%%%%%%%%%%%%%%%%%%%%%%%%%%%%%%%%%%%%%%%%%%%%%%%%%%%%%%%%%

\subsection{Modified Rates in SM Higgs Search Channels}
We define the ratios of the Higgs branchings to photons and to SM fermions, $f$, or gauge bosons, $V=W,~Z$, to their values in the SM as
\begin{align}
&{\cal B}\left(h\to\gamma\gamma\right)_{\rm eff}=R_{\gamma\gamma}\times{\cal B}_{\rm SM}\left(h\to\gamma\gamma\right),
\label{eq:Rgaga}
\\
&{\cal B}\left(h\to f\bar f,VV\right)=R_{XX}\times{\cal B}_{\rm SM}\left(h\to f\bar f,VV\right).
\label{brs}
\end{align}
The presence of the $h\rightarrow aa$ decay suppresses uniformly the rates into all SM final states, yielding
\begin{align}
R_{XX}=1-{\cal B}(h\rightarrow aa).
\label{eq:suppr}
\end{align}
Note that to avoid cluttering notation, and because the rescaling is the same in both cases, we use $R_{XX}$ to represent both the fermionic and $W,Z$ rescalings relative to the SM.

The suppression in Eq.~\ref{eq:suppr} is also present for the pure $\gamma\gamma$ final state; however, the subscript ``${\rm eff}$" in Eq.~(\ref{eq:Rgaga}) indicates that in pseudoscalar models there can be an additional, effective contribution to the measured $h\rightarrow\gamma\gamma$ rate. Since the LHC detectors have finite resolution, a fraction of $h\to aa\to 4\gamma$ events may have sufficiently boosted photon pairs that each pair is identified as a single photon. Assuming a SM production cross section for $h$ and a 100\% branching of $a\rightarrow\gamma\gamma$, the measured diphoton branching ratio will be
\begin{align}
{\cal B}\left(h\to\gamma\gamma\right)_{\rm eff}&={\cal B}\left(h\to\gamma\gamma\right)+\epsilon\times{\cal B}\left(h\to aa\right)
\end{align}
or
\begin{align}
R_{\gamma\gamma}&=1+{\cal B}(h\rightarrow aa)\left(\frac{\epsilon}{{{\cal B}_{\rm SM}\left(h\to\gamma\gamma\right)}}-1\right).
\label{Rdef}
\end{align}
In these formulae, $\epsilon$ is the probability that both $\gamma\gamma$ pairs are identified as single photons, so that four photons appear as two. We see that to achieve an effective diphoton rate greater than or equal to the SM rate requires 
\begin{align}
\epsilon\ge{\cal B}_{\rm SM}\left(h\to\gamma\gamma\right)\simeq 0.0023
\end{align} 
for $m_h=125\GeV$.

%%%%%%%%%%%%%%%%%%%%%%%%%%%%%%%%%%%%%%%%%%%%%%%%%%%%%%%%%%%%%%%%%%%%%%%%%%%%%%%%

\subsection{$4\gamma\rightarrow 2\gamma$ Misidentification Rate}
\label{sec:misid}
Estimating $\epsilon$ is complicated by several factors. We would like to be conservative in our estimate of the expected rate; on the other hand, underestimating the rate by too much might falsely indicate that some model points are allowed, when in fact the true rates are so large that the points are already ruled out by the LHC. We base our estimate of $\epsilon$ on the ATLAS selection criteria used to identify isolated photons in their cut-based analysis, and we comment on differences with CMS. 

ATLAS uses a number of calorimeter variables to parametrize the shape of an electromagnetic shower, which can then be used to discriminate true isolated photons from backgrounds. The background from isolated $\pi^0\rightarrow\gamma\gamma$ decays bears strong similarity to our $a\rightarrow\gamma\gamma$ process, and ATLAS efficiently vetoes isolated pions using information from the first layer of the calorimeter, which has finely-segmented strips in $\eta$~\cite{Aad:2009wy}. The primary discrimination variables in the first layer are ${\bf E_{\rm ratio}}$, which is the difference in energies between two strips containing energy maxima normalized to their sum; ${\bf \Delta E}$, which measures the difference in energies between the strip with the second-largest energy maximum and the strip with the minimum energy between the first two maxima; ${\bf F_{\rm side}}$, which is the fraction of the energy deposited in seven strips in $\eta$ around the maximum that does not fall into the central three strips; ${\bf w_{s \rm 3}}$, which measures the energy deposition in the two strips adjacent in $\eta$ to a strip with an energy peak, relative to the total energy in the three strips; and ${\bf w_{s \rm tot}}$, which generalizes $w_{s3}$ to approximately twenty strips in $\eta$ and two strips in $\phi$~\cite{Aad:2010sp}.

To simplify our analysis, we begin by restricting our attention to photons that do not convert to $e^+e^-$ pairs in the tracker, and which are so highly collimated that a second energy maximum does not appear in the first-layer calorimeter strips. In this case $\Delta E$ is set to 0, $E_{\rm ratio}$ is set to 1, and most of the energy will be deposited into just a few adjacent strips. Therefore we expect that the most sensitive variable will be $w_{s3}$, defined precisely as
\begin{align}
w_{s3}\equiv\sqrt{\sum_i E_i(i-i_{\rm max})^2/\sum_iE_i},
\end{align}
where $i$ labels the strips.  On average the photon pair from a $\pi^0$ decay generates a larger $w_{s3}$ value than a true single-photon event and thus may be efficiently rejected. On the other hand, the probability that a photon pair passes the $w_{s3}$ cut should increase with the collimation of the pair. To easily estimate $\epsilon$, we would like to approximate the cut on $w_{s3}$ by a cut on the photon pair separation.

For unconverted events, ATLAS uses a weakly $\eta$-dependent cut on $w_{s3}$ that is typically between $0.6-0.7$, and is about $0.66$ for the most central strips in the barrel. The average value of $w_{s3}$ for true photons in these strips is about 0.58~\cite{Collaboration:1345329}.   We can reproduce this number with the following simple model.  First, we assume that a single photon lands in the center of one of these most central strips, and deposits its energy according to a Gaussian distribution in $\eta$ with standard deviation $0.52$ times the smallest strip width. This value is chosen so that the photon deposits about 70\% of the energy into the central strip and 15\% into each adjacent strip, giving a $w_{s3}$ value of 0.58.  Subsequently, we compute $w_{s3}$ for a pair of photons as a function of $\eta$ separation, averaging over the impact point in the central strip. We find that for $\Delta\eta_{\gamma\gamma}=0.0015$, $w_{s3}\simeq 0.66$, equal to the ATLAS cut in these strips. On the other hand, the central strips have a width in $\eta$ of $\Delta\eta_{\rm strip}=0.0031$ (the smallest in the calorimeter.)  Therefore, we estimate that the cut on $w_{s3}$ can be simulated by a cut on the photon separation, given by
\begin{align}
\Delta\eta_{\gamma\gamma}<1/2\times\Delta\eta_{\rm{strip}}.
\label{eq:etacut}
\end{align}
For larger $\eta$, where larger strips are present, we also require a separation less than $1/2$ of the relevant strip size, which is probably mildly conservative since on average the energy leakage into adjacent strips from a single-photon hit will be less.

There are two simple ways in which $w_{s3}$ may become insufficient. First, photons that are less highly collimated (such as those that appear in the decays of more massive pseudoscalars) may deposit energy in strips that are sufficiently separated (more than three strips apart) that $w_{s3}$ becomes insensitive. We assume that such events are very efficiently rejected by the other first-layer discriminators described above, and therefore our cut on $\Delta\eta_{\gamma\gamma}$ is still a good proxy for the cuts on those variables. Secondly, photons may be closely spaced in $\eta$, but more broadly spaced in $\phi$. Since the segmentation in $\phi$ is much coarser than in $\eta$, such events must be very separated in $\phi$ in order for cuts on the variable $R_{\phi}$ (which measures the energy distribution across several second-layer $\phi$ cells) to reject them. A cut on $\Delta\phi_{\gamma\gamma}$ can simulate the cut on $R_{\phi}$, but since such events are geometrically rare, the net efficiency is quite insensitive to the precise value of this cut. For definiteness we set the cut on $\Delta\phi_{\gamma\gamma}$ to be equal to the smallest second-layer cell size in $\phi$ as a function of $\eta$ ($\Delta\phi_{\rm{cell}}=0.025$ for $\eta<1.4$):
\begin{align}
\Delta\phi_{\gamma\gamma}<\Delta\phi_{\rm{cell}}.
\label{eq:phicut}
\end{align}

So far we have discussed only events which contain no converted photons. Conversions happen with an $\eta$- and $E_T$-dependent probability that ranges from about 10\% at low $\eta$ to more than 50\% at larger $\eta$~\cite{Aad:2010sp}. Since $h\rightarrow aa$ events produce four photons in the final state instead of two, a larger fraction of our events will contain at least one converted photon than are contained in pure $h\rightarrow\gamma\gamma$ events.  In our conversion events, one cluster in the calorimeter may contain one $e^+e^-$ pair and one $\gamma$, or two $e^+e^-$ pairs, so we might imagine that these events would be easy to distinguish from pure single-$\gamma$ events and may even be vetoed. In the case with $\gamma e^+e^-$, there is a mismatch between the momentum measured by the tracker (sensitive only to the charged particles) and the energy deposit in the calorimeter. In the case with $2e^+e^-$, two conversion vertices may be reconstructed in the tracker. Both of these signatures have been considered for rejecting isolated $\pi^0$  backgrounds~\cite{oldtdr}; however, at present, the ATLAS photon identification analysis does not use an $E/p$ cut, and does not automatically veto events with multiple conversion vertices~\cite{Aad:2010sp} (only the ``best" reconstructed conversion vertex is used, and is determined by the conversion radius and the number of tracks associated with the vertex.) Furthermore, the cuts on the calorimeter variables are relaxed somewhat to accommodate the fact that the energy deposits for single-photon conversions tend to spread mildly in $\eta$ and considerably in $\phi$ (due to the magnetic field.) Therefore, we will make the approximation that the value of $\epsilon$ relevant for $4\gamma$ events containing conversions is the same as the value of $\epsilon$ for the unconverted sample; namely, it is determined by applying the collimation cuts in Eqs.~(\ref{eq:etacut},\ref{eq:phicut}) to the sample of parent photons before they convert.

In summary, for all classes of photon events, we make the approximation
\begin{align}
\epsilon\simeq \epsilon_{\rm coll},
\label{epsapprox}
\end{align}
where $\epsilon_{\rm coll}$ is the rate at which the $4\gamma$ events are expected to pass the $\eta$-dependent collimation cuts $\Delta\eta_{\gamma\gamma}<1/2\times\Delta\eta_{\rm{strip}}$, $\Delta\phi_{\gamma\gamma}<\Delta\phi_{\rm{cell}}$.

The collimation of a photon pair produced in an $a$ decay is controlled by the ratio $2m_a/E_a$, where $E_a$ is determined by the momentum of the Higgs boson. We simulate $7$ TeV $gg\rightarrow h\rightarrow aa\rightarrow 4\gamma$ events in Madgraph 5~\cite{Alwall:2011uj}, and on the subsample of events with all photons satisfying $|\eta|\in(0,1.37)\cup(1.52,2.37)$ (the region in which ATLAS reconstructs photon candidates for the $h\rightarrow\gamma\gamma$ analysis), and which pass the ATLAS photon $p_T$ cuts~\cite{ATLAS:2012ad}, we compute the efficiency $\epsilon_{\rm coll}$. In Fig.~\ref{fig:eff} we plot $\epsilon_{\rm coll}$ as a function of $m_a$.\footnote{Production mechanisms other than gluon fusion result in somewhat different kinematic distributions for $h$. However, gluon fusion dominates the production with O(10\%) contribution from all other channels~\cite{ATLAS:2012ad}, so the effects on $\epsilon_{\rm coll}$ from these processes will be second-order. We neglect them in this study. In addition, we have checked that the effect of raising $\sqrt{s}$ to $8$ TeV is negligible.}

\begin{figure}
\rotatebox{0}{\resizebox{80mm}{!}{\includegraphics{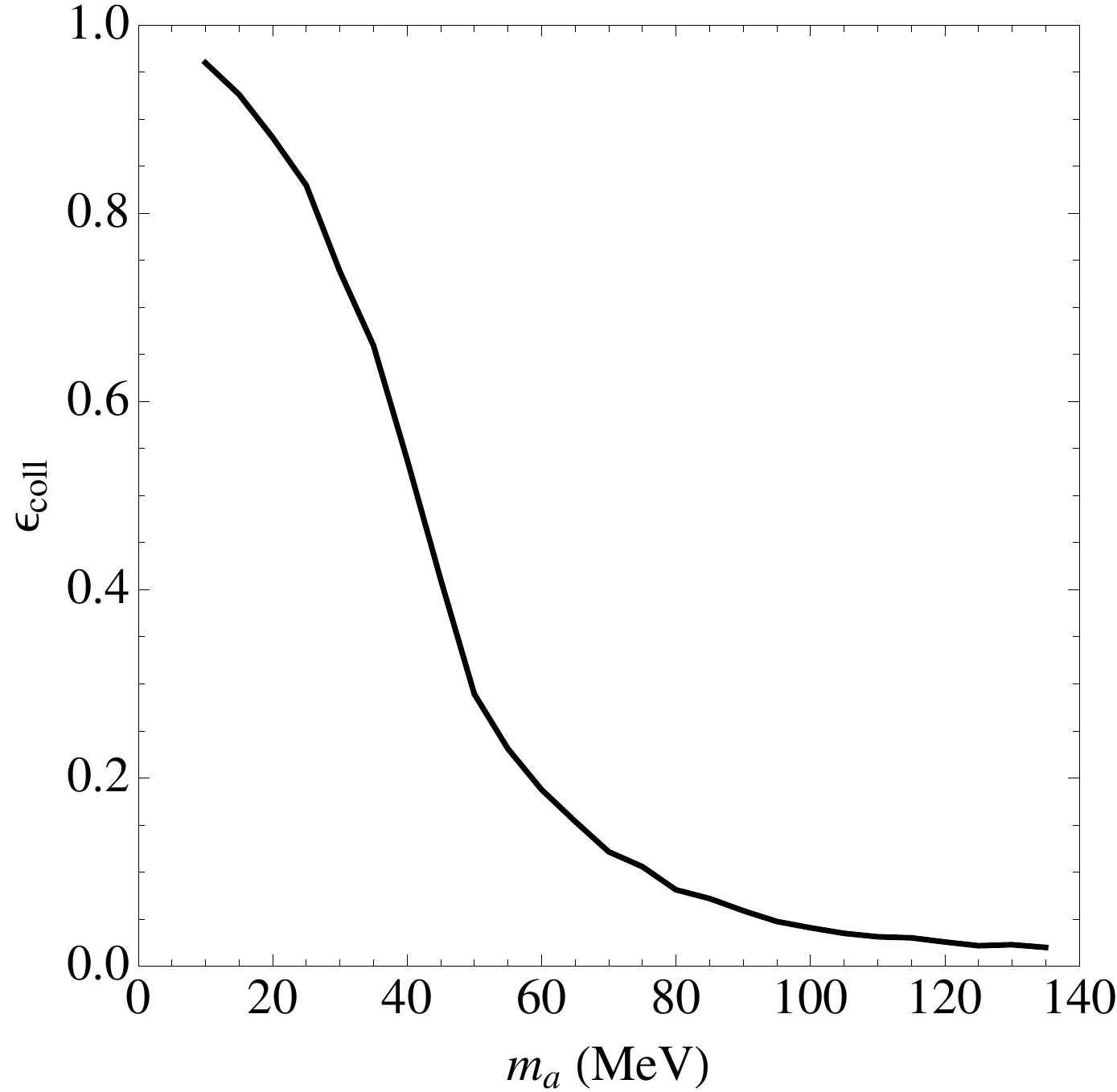}}}
\caption{Fraction of events where both photon pairs are sufficiently collimated to pass ATLAS isolation cuts, as a function of the pseudoscalar mass.}
\label{fig:eff}
\end{figure}

To cross-check that our cuts effectively reproduce the rejection power of $w_{s3}$ and the other discrimination variables, we use the known isolated $\pi^0$ rejection power of ATLAS as a function of $(E_T,\eta)$ (given in Fig. 11 of Sec. 5 of~\cite{Aad:2009wy}) to estimate $\epsilon$ from our simulation at the mass point $m_a=m_{\pi^0}$.  We find good agreement between the $\epsilon$ derived from this method and $\epsilon_{\rm coll}$ obtained from the collimation cuts. Since the ATLAS $\pi^0$ rejection power was defined on events containing both converted and unconverted photons, we are reassured that Eq.~(\ref{epsapprox}) is plausible.

Unlike the ATLAS analysis, the CMS analysis does place cuts on the ratio of the calorimeter energy measurement to the momentum measured in the tracker in order to isolate single photons from $\pi^0$ decays~\cite{CMS-PAS-EGM-10-006}. The cut may reduce $\epsilon$ relative to $\epsilon_{\rm coll}$. However, the CMS ECAL has a barrel granularity $\simeq6\times$ larger in $\eta$ than that of ATLAS~\cite{Ball:2007zza}, increasing $\epsilon_{\rm coll}$. For this analysis we assume that the net efficiencies $\epsilon$ will be comparable between the two detectors, but in principle any difference between them translates into an experiment-dependent and possibly conversion-dependent prediction for the $\gamma\gamma$ rate and could be used to probe the value of $m_a$.

To summarize this section: 

(1) High collimation is sufficient for a significant number of $2\gamma$ clusters to be misidentified as single isolated photons by the experimental analyses.

(2) In order to produce an interesting misidentification rate, $m_a\lesssim m_{\pi^0}$. For larger $m_a$ there may still be misidentified events in the $h\rightarrow\gamma\gamma$ signal, particularly if ${\cal B}(h\rightarrow aa)$ is not small, but the scenario should be more properly studied in the context of an $h\rightarrow 4\gamma$ signal. 

(3) Contamination of $h\rightarrow\gamma\gamma$ by misidentified $4\gamma$ events will be visible as an excess of the total number of conversion events relative to the $\gamma\gamma$ expectation, as an excess of conversion events with multiple reconstructed vertices, and as an excess of conversion events with a mismatch between the track $p_T$ and the energy deposit in the ECAL.

%%%%%%%%%%%%%%%%%%%%%%%%%%%%%%%%%%%%%%%%%%%%%%%%%%%%%%%%%%%%%%%%%%%%%%%%%%%%%%%%

\subsection{Current Constraints and Fits}
In this study we are interested in the possibility that the current excesses in Higgs searches at ATLAS, CMS, and the Tevatron are due to a Higgs boson with mass near 125 GeV. Under this hypothesis, some parts of the $(m_a,{\cal B}(h\rightarrow aa))$ parameter space are disfavored: they produce too few vector boson or fermion events, or too many or too few diphoton-like events.  We estimate the constraints with a matched-filter, taking each point in parameter space as a template and estimating a best-fit amplitude $\hat{R}$ for the signal strength at the point.\footnote{For other recent studies of the Higgs best-fit cross section data using similar $\chi^2$ analyses, see~\cite{Carmi:2012yp,*Espinosa:2012ir,*Giardino:2012ww,*Azatov:2012bz}.} In the Gaussian limit $\hat{R}$ can be estimated as
\begin{align}
\hat{R}=\sigma^2 t_i C^{-1}_{ij} d_j ,
\end{align}
where $d_j$ is the ``data," which we take to be the best-fit amplitudes relative to the SM for the channels 
$h\rightarrow\gamma\gamma$~\cite{ATLAS:2012ad}, 
$h\rightarrow ZZ\rightarrow 4l$~\cite{ATLAS:2012ac}, 
$h\rightarrow WW\rightarrow l\nu l\nu$~\cite{ATLAS-CONF-2012-012}, 
$(W/Z)h\rightarrow (ll,l\nu,\nu\nu)b\bar{b}$~\cite{ATLAS-CONF-2012-015}, 
and 
$h\rightarrow \tau\tau$~\cite{ATLAS-CONF-2012-014}, 
presented by ATLAS at $m_h=126$ GeV~\cite{ATLAS-CONF-2012-019} , the channels 
$gg\rightarrow h\rightarrow\gamma\gamma$~\cite{Chatrchyan:2012tw,CMS-PAS-HIG-12-001}, 
$qqh\rightarrow qq\gamma\gamma$~\cite{Chatrchyan:2012tw,CMS-PAS-HIG-12-001}, 
$h\rightarrow ZZ\rightarrow 4l$~\cite{Chatrchyan:2012dg}, 
$h\rightarrow WW\rightarrow l\nu l\nu$~\cite{Chatrchyan:2012ty}, 
$(W/Z)h\rightarrow (ll,l\nu,\nu\nu)b\bar{b}$~\cite{Chatrchyan:2012ww}, 
and 
$h\rightarrow \tau\tau$~\cite{Chatrchyan:2012vp}, 
presented by CMS at $m_h=124$ GeV~\cite{Chatrchyan:2012tx}, and the channels 
$h\rightarrow b\bar{b}$ 
and 
$h\rightarrow WW$ 
presented by CDF and DZero at $m_h=125$ GeV~\cite{TEVNPH:2012ab}. For ATLAS and CMS we chose slightly different mass points based on where their respective $\gamma\gamma$ excesses are most significant; we assume that the current experimental resolution is large enough that both excesses can come from the same Higgs-like particle. The vector $t_i$ gives the theoretical prediction for each data point (either $R_{\gamma\gamma}$ or $R_{XX}$), and $C^{-1}_{ij}$ is the inverse covariance matrix set by the squared symmetrized error bars (we include also correlations which we estimate from the luminosity uncertainty and the theoretical uncertainty on the gluon fusion cross section). The error on the estimate $\hat{R}$ is given by
\begin{align}
\sigma \equiv ( t_i C^{-1}_{ij} t_j )^{-1/2}.
\end{align}

\begin{figure}
\rotatebox{0}{\resizebox{80mm}{!}{\includegraphics{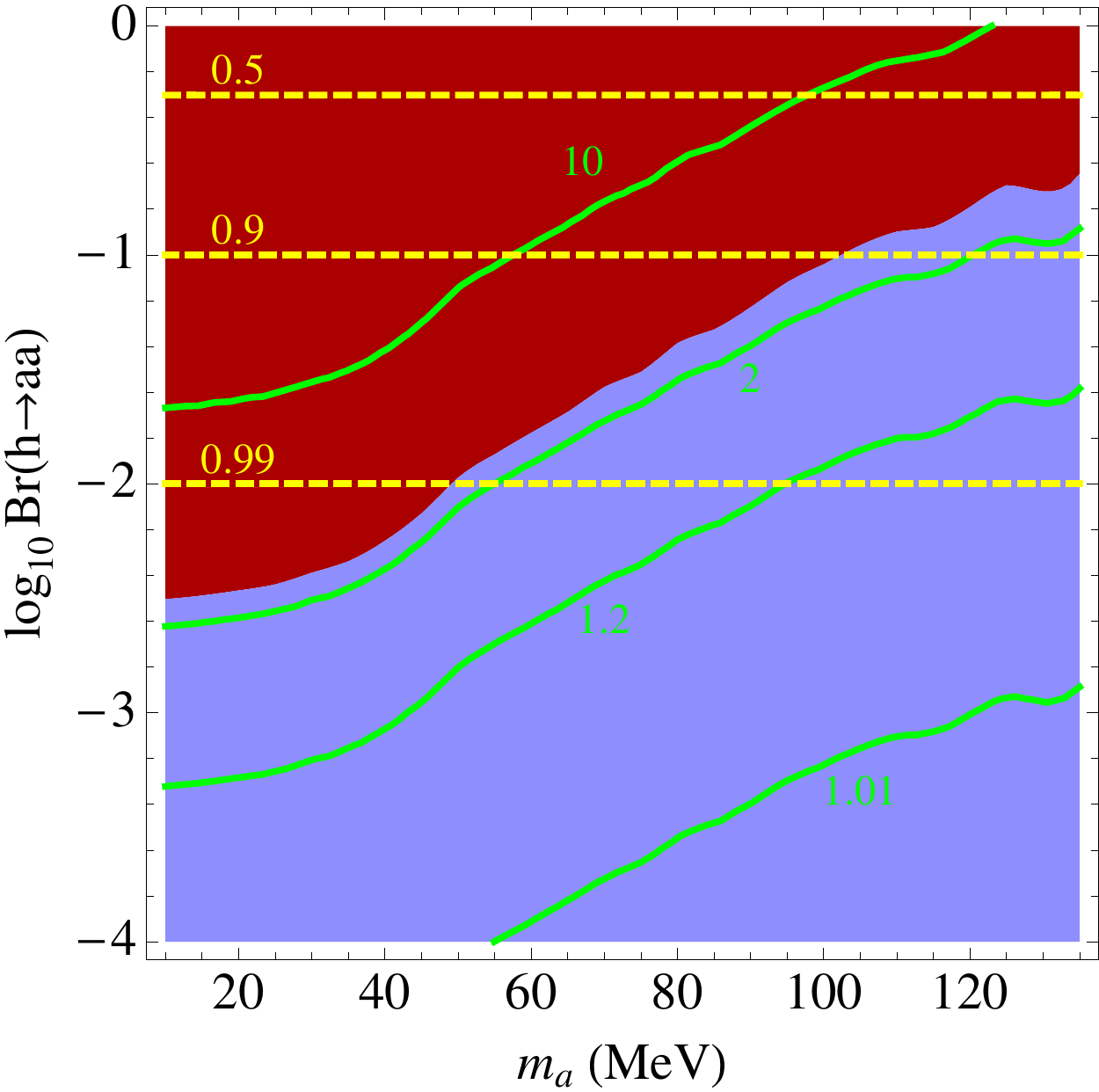}}}
\caption{Regions of light pseudoscalar parameter space that are favored (blue/light) and disfavored (red/dark) by the current best-fit signal strengths in the $h\rightarrow\gamma\gamma,ZZ,WW,bb,\tau\tau$ channels. Contours are overlaid for the net diphoton (solid green) and $ZZ,WW,bb,\tau\tau$ rates (dashed yellow) expected at the LHC relative to the SM rates.}
\label{fig:excl}
\end{figure}

Model points where $R=1$ is outside the 90\% CL band around $\hat{R}$ are then disfavored. Of course, this statistical procedure is quite approximate at this stage, and serves only to get an estimated exclusion region if the data sets are large and $m_h$ actually is at 125 GeV. A precise calculation would construct the full Poisson likelihood with all correlations. With these caveats in mind, in Fig.~\ref{fig:excl} we present the regions of $(m_a,{\cal B}(h\rightarrow aa))$ parameter space disfavored by the current LHC and Tevatron results, including contours of $R_{\gamma\gamma}$ and $R_{XX}$. The contours make clear the fact that the disfavored regions are mainly controlled by the excesses in the diphoton channels: increasing it above a few times the SM rate is in tension with the current excesses. We also maximize the likelihood over the plane, fixing the signal strength parameter to $R=1$ and constraining $m_a<m_\pi$. The best-fit point lies near the $R_{\gamma\gamma}=2$ contour at the $m_\pi$ boundary; relaxing the constraint, it would move beyond to near the intersection of the $R_{\gamma\gamma}=2$ and $R_{XX}=0.5$ contours. However, the data are over-fit, $\chi^2/{\rm d.o.f.}=7.3/10$, and the statistic is very shallow, particularly in the direction of constant $R_{\gamma\gamma}$. Therefore, the precise location of the best-fit point has little meaning, and we do not show it in Fig.~\ref{fig:excl}. For comparison, we find that the SM $\chi^2/{\rm d.o.f.}=12.2/12$.

\end{section}

%%%%%%%%%%%%%%%%%%%%%%%%%%%%%%%%%%%%%%%%%%%%%%%%%%%%%%%%%%%%%%%%%%%%%%%%%%%%%%%%
%%%%%%%%%%%%%%%%%%%%%%%%%%%%%%%%%%%%%%%%%%%%%%%%%%%%%%%%%%%%%%%%%%%%%%%%%%%%%%%%
%%%%%%%%%%%%%%%%%%%%%%%%%%%%%%%%%%%%%%%%%%%%%%%%%%%%%%%%%%%%%%%%%%%%%%%%%%%%%%%%

\begin{section}{Direct Constraints on $a$}
\label{sec:directconstr}
We now discuss existing experimental constraints on the model in Eq.~(\ref{eq:L_int}).  Similar considerations have been undertaken with light pseudoscalars in~\cite{Masso:1995tw,*Redondo:2008tq,Andreas:2010ms}.

Constraints from $e^+e^-\rightarrow\gamma a$, quarkonia decays, and beam dump experiments are robust in the sense that they are controlled largely or entirely by the coupling of $a$ to photons. Constraints from the muon anomalous magnetic moment and meson decays involving flavor-changing neutral currents are sensitive not only to the $aF\tilde{F}$ coupling, but also any small coupling of $a$ to SM fermions that might be present in the Lagrangian. These contributions may interfere constructively or destructively with the $aF\tilde{F}$ terms-- we can even envision situations where some constraints are mitigated by finely tuned tree level couplings of $a$ to SM fermions. We assume for the purpose of this section that the fermionic couplings at the scale $M$ are small enough that the leading contributions to the constraints come from photon loops. 

Since $aF\tilde F$ is dimension-5, the amplitudes are typically logarithmically divergent. To remove the divergence, we cut the integrals off at momenta $\sim M$, which is motivated by the view that the $aF\tilde F$ interaction is an effective coupling resulting from some new physics at a scale $M$---whether $a$ is a composite with constituents with masses of that order or is fundamental and coupled to photons via particles with such masses.  The phenomenological need for $M$ to be tens of GeV poses some model-building puzzles in this respect and we will sketch a few potential scenarios leading to such a scale in Sec.~\ref{sec:models}.  

Taken together, these low energy constraints should be viewed mainly illustratively.  We attempt to be as conservative as possible, erring on the side of presenting stronger limits on the parameter space of the model and comment that these limits can likely be circumvented with relatively straightforward extensions of the model under consideration.

%%%%%%%%%%%%%%%%%%%%%%%%%%%%%%%%%%%%%%%%%%%%%%%%%%%%%%%%%%%%%%%%%%%%%%%%%%%%%%%%

\begin{subsection}{Muon Anomalous Magnetic Moment}
\label{sec:g-2}
A light pseudoscalar that couples to two photons will contribute to the anomalous magnetic moment of the muon.  The leading correction, shown in Fig.~\ref{fig:g-2}, occurs at order $\alpha^3$ and is analogous to the $\pi^0$-pole portion of the hadronic light-by-light contribution to $(g-2)_\mu$.
\begin{figure}
\rotatebox{0}{\resizebox{60mm}{!}{\includegraphics{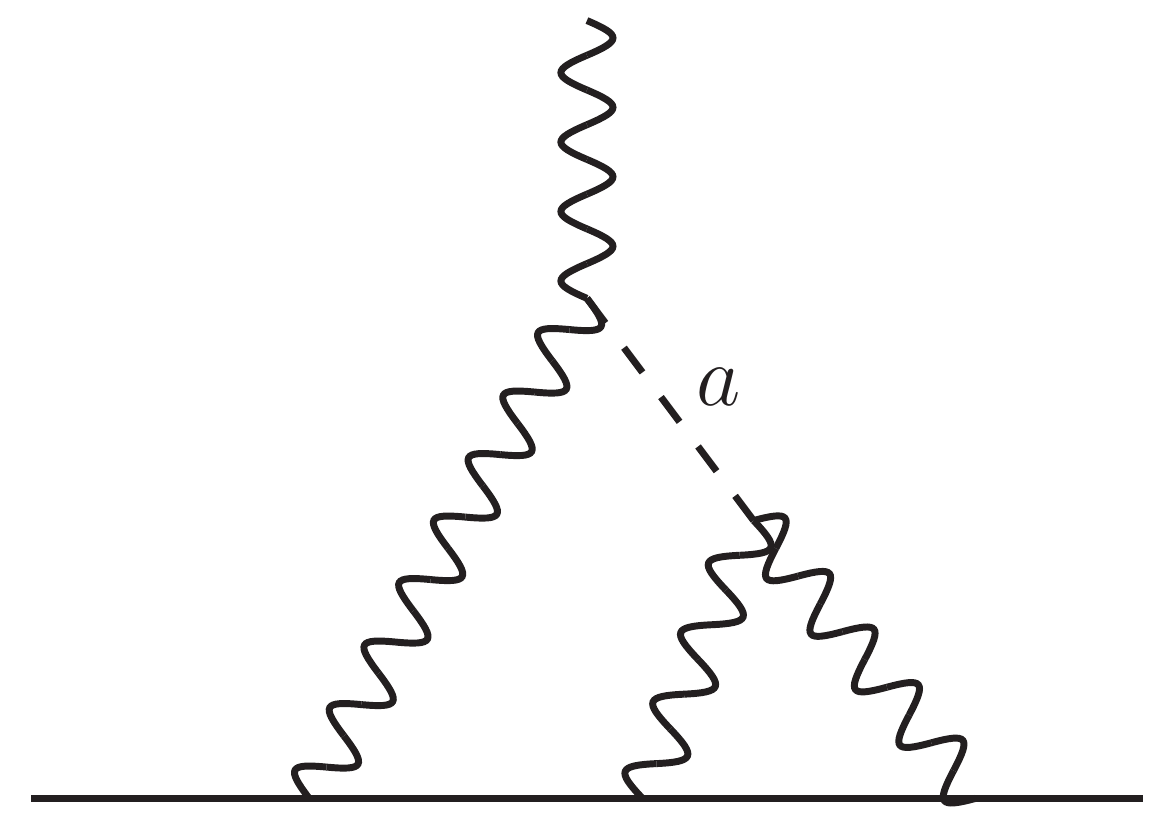}}}
\caption{A representative diagram of the leading contribution of the pseudoscalar, $a$, to $(g-2)_\mu$.}
\label{fig:g-2}
\end{figure}

To estimate this effect as a function of $m_a$ and $M$, we make use of the expression for the $\pi^0$-pole contribution to $(g-2)_\mu$ in~\cite{Knecht:2001qf}, rescale it by the strength of the photon coupling relative to that of $\pi^0$, $\left(4\pi^2 F_\pi/M\right)^2$, replace the pion mass with $m_a$, and cut off the logarithmic divergence at the scale $M$ instead of at $\sim 4\pi^2 F_\pi$.  To set limits we require that this contribution is less than the current deviation between experiment and theory on $\Delta a_\mu=(g-2)_\mu/2$ of about $25\times 10^{-10}$~\cite{Teubner:2010ah,*Davier:2010nc}.  This sets a lower limit on $M$ around $4\pi^2 F_\pi=3.64\GeV$ that is not very sensitive to $m_a$ for the range of pseudoscalar masses we consider.

Because of the lepton mass dependence of corrections to $g-2$, the electron anomalous magnetic moment is less constraining and does not appear in our limits.
\end{subsection}

%%%%%%%%%%%%%%%%%%%%%%%%%%%%%%%%%%%%%%%%%%%%%%%%%%%%%%%%%%%%%%%%%%%%%%%%%%%%%%%%

\begin{subsection}{$e^+e^-\to\gamma a$ Limits}
\label{sec:ee}
The reaction $e^+e^-\to\gamma a$ can proceed through the $s$-channel exchange of a virtual photon.  The cross section for this process is independent of the center-of-mass energy,
\begin{align}
\frac{d\sigma}{d\cos\theta}=\frac{2\pi^2\alpha^3}{3M^2}\,\frac{3\left(1+\cos^2\theta\right)}{8},
\end{align}
where $\theta$ is the angle between the photon and the beam axis in the center-of-mass.  If $a$ decays without being detected, then this process is subject to limits from $e^+e^-\to\gamma+{\rm inv}$.  We apply limits from the DELPHI collaboration~\cite{Abdallah:2003np,*Abdallah:2008aa} using $650~{\rm pb}^{-1}$ of data at center-of-mass energies $\sqrt{s}=180\GeV$--$209\GeV$.\footnote{We note here that limit on $e^+e^-\to\gamma+{\rm inv.}$ from the ASP experiment~\cite{Hearty:1989pq} that has been used in previous studies to constrain light pseudoscalars coupled to photons do not actually apply.  This is because their analysis imposed a cut on the photon's energy of $E_\gamma<\sqrt{s}/2-4.5\GeV=10\GeV$ to eliminate backgrounds from $e^+e^-\to\gamma\gamma$, rendering it insensitive to invisibly decaying particles produced in association with a photon with a mass less than $16\GeV$.}  In the photon energy range of interest for $m_a\ll \sqrt{s}$, $E_\gamma\simeq\sqrt{s}/2$, we require that there are fewer than 20 events and assume that the product of angular acceptance and efficiency is $\sim 0.5$ .  Assuming that the $a$ decay length must be larger than 2~m to be unseen at DELPHI and a boost of $\gamma\sim 90\GeV/m_a$ {\em excludes} the following region of parameter space:
\begin{align}
110\GeV\left(\frac{m_a}{40\MeV}\right)^2\lsim M\lsim125\GeV.
\end{align}
\end{subsection}

%%%%%%%%%%%%%%%%%%%%%%%%%%%%%%%%%%%%%%%%%%%%%%%%%%%%%%%%%%%%%%%%%%%%%%%%%%%%%%%%

\begin{subsection}{Meson Decays}
\label{sec:mesons}
The interactions in Eq.~(\ref{eq:L_int}) lead to effective couplings of $a$ to SM fermions at the one-loop level.  This can have effects on rare meson decays involving photons or missing energy if $a$ decays too late to be detected.

Effective couplings to up-type quarks can then lead to flavor-changing transitions in the down sector such as $s\to d+a$ as in Fig.~\ref{fig:stod}.
\begin{figure}
\rotatebox{0}{\resizebox{60mm}{!}{\includegraphics{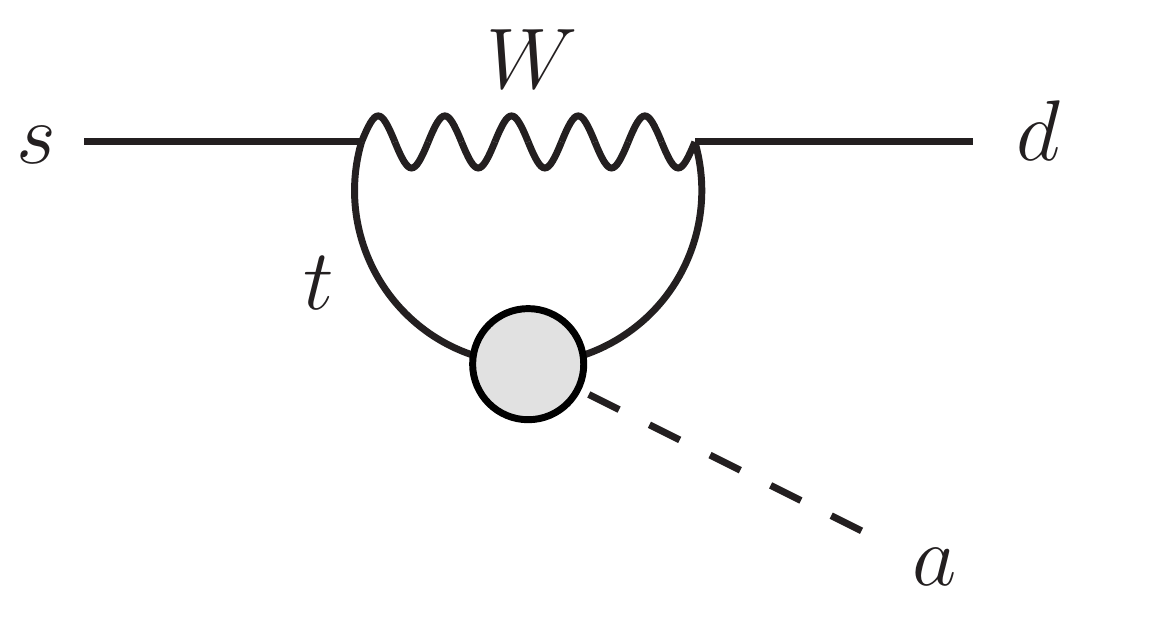}}}
\caption{Diagram that gives the leading contribution to $s\to d+a$ from an effective interaction between $a$ and the top quark.}
\label{fig:stod}
\end{figure}
This leads to the flavor-changing decay $K^\pm\to\pi^\pm a$ which must be confronted with experimental data on $K^\pm\to\pi^\pm\gamma\gamma$.  The amplitude for $s\to d+a$ can be very roughly estimated as
\begin{align}
{\cal M}\left(s\to d+a\right)&\sim\frac{\alpha^2}{M}\log^2\left(\frac{M^2}{m_t^2}\right)\left(\frac{G_F m_t^2}{4\sqrt{2}\pi^2}\right)
\\
&~~~~\times V_{td}^\ast V_{ts} \left(m_s\bar d_L s_R-m_d\bar d_R s_L\right),
\nonumber
\end{align}
where we have again cut off the logarithmic divergence at the scale $M$.  Using this we arrive at a rate for $K^\pm\to\pi^\pm a$ of
\begin{align}
&\Gamma\left(K^\pm\to\pi^\pm a\right)\sim \frac{\alpha^4m_s^2}{M^2}\log^4\left(\frac{M^2}{m_t^2}\right)\left(\frac{G_F^2 m_t^4}{2^{10}\pi^5}\right)
\\
&~~~~~~~~\times\frac{p_\pi}{m_K^2}\left|V_{td}^\ast V_{ts}\right|^2\left(\frac{m_K^2-m_\pi^2}{m_s-m_d}\right)^2\left|f_0^{K\pi}\left(0\right)\right|^2,
\nonumber
\end{align}
where the form factor $f_0^{K\pi}\left(0\right)\simeq 1$~\cite{Marciano:1996wy} and $p_\pi$ is the pion momentum in the kaon rest frame.   We apply the limits on hypothetical particles with masses less than $100\MeV$ decaying to two photons from~\cite{Kitching:1997zj} (generally at the $10^{-7}$ level) to ${\cal B}\left(K^+\to\pi^+a\right)$, resulting in $M\gsim 1$--$2\GeV$.  In addition, if $a$ is sufficiently long-lived so that its decay is not detected, it will give a contribution to the process $K^+\to\pi^++{\rm inv}$.  The current experimental upper limit on the branching ratio for $K^+$ to decay to $\pi^+$ and a light invisible particle is $7.3\times10^{-11}$~\cite{Anisimovsky:2004hr}.  For this limit to apply, $a$ must have a decay length larger than about 1~m when produced with a $\pi^+$ in the decay of a $K^+$ at rest.  The {\em excluded} region is then
\begin{align}
1500\GeV\left(\frac{m_a}{40\MeV}\right)^2\lsim M\lsim 17\GeV.
\end{align}
Analogous limits from $B^\pm\to K^\pm$ transitions are less stringent in the pseudoscalar mass range of interest.

$^3{\rm S}_1$ quarkonium states can decay to $\gamma a$ through an $s$-channel virtual photon, just as in $e^+e^-$ annihilation.  In the $\bar cc$ and $\bar bb$ systems, assuming $m_a\ll m_{c,b}$, the branching ratios are
\begin{align}
{\cal B}\left(J/\psi\to\gamma a\right)&\simeq\frac{2\pi\alpha m_c^2}{M^2}{\cal B}\left(J/\psi\to e^+e^-\right)
\nonumber
\\
&\simeq4.6\times10^{-6}\left(\frac{10\GeV}{M}\right)^2,
\\
{\cal B}\left(\Upsilon(1{\rm S})\to\gamma a\right)&\simeq\frac{2\pi\alpha m_b^2}{M^2}{\cal B}\left(\Upsilon(1{\rm S})\to e^+e^-\right)
\nonumber
\\
&\simeq2.3\times10^{-5}\left(\frac{10\GeV}{M}\right)^2.
\end{align}
We require that the contribution to $J/\psi\to 3\gamma$ from $J/\psi\to\gamma a$, $a\to \gamma\gamma$ is less than the experimental result ${\cal B}\left(J/\psi\to 3\gamma\right)=\left(1.2\pm0.4\right)\times10^{-5}$~\cite{Adams:2008aa}.  This translates into a limit $M\gsim6.2\GeV$.  Current limits on ${\cal B}\left(\Upsilon(1{\rm S})\to\gamma a\right)$ at the $10^{-6}$ to $10^{-5}$ level assume $a$ decays to leptons or hadrons~\cite{Love:2008aa,*McKeen:2008gd} and therefore are not constraining on a pseudoscalar that decays promptly to $\gamma\gamma$.  In addition, if $a$ is long-lived enough to decay after passing through the detector these decays are also subject to limits on $J/\psi,\, \Upsilon(1{\rm S})\to\gamma+{\rm inv}$ as in the $e^+e^-\to\gamma+{\rm inv.}$ case in Sec.~\ref{sec:ee}.  The current experimental limits are ${\cal B}\left(J/\psi\to\gamma a\to\gamma+{\rm inv.}\right)<4.6\times10^{-6}$ for $m_a<150\MeV$~\cite{Insler:2010jw} and ${\cal B}\left(\Upsilon(1{\rm S})\to\gamma a\to\gamma+{\rm inv.}\right)<1.4\times10^{-5}$ for $m_a<5\GeV$~\cite{Balest:1994ch}.  To estimate the region of parameter space that these limits are sensitive to, we assume that the $a$ decay length needs to be greater than 1~m to go undetected in these searches.  This {\em excludes} the regions
\begin{align}
&590\GeV\left(\frac{m_a}{40\MeV}\right)^2\lsim M\lsim10\GeV,
\end{align}
from $J/\psi\to\gamma+{\rm inv.}$ and
\begin{align}
&330\GeV\left(\frac{m_a}{40\MeV}\right)^2\lsim M\lsim13\GeV,
\end{align}
from $\Upsilon(1{\rm S})\to\gamma+{\rm inv}$.

%We note that the limits from $K^+\to\pi^++{\rm inv}$ and $J/\psi,\, \Upsilon(1{\rm S})\to\gamma+{\rm inv.}$ are less stringent than those from $e^+e^-\to\gamma+{\rm inv.}$ in Sec.~\ref{sec:ee}.

We note that the regions of $\left(m_a,M\right)$ ruled out by $K^+\to\pi^++{\rm inv}$ and $J/\psi,\, \Upsilon(1{\rm S})\to\gamma+{\rm inv.}$ are all contained within the region excluded by $e^+e^-\to\gamma+{\rm inv.}$ in Sec.~\ref{sec:ee}.

\end{subsection}

%%%%%%%%%%%%%%%%%%%%%%%%%%%%%%%%%%%%%%%%%%%%%%%%%%%%%%%%%%%%%%%%%%%%%%%%%%%%%%%%

\begin{subsection}{Beam Dump Experiments}
\label{sec:beamdump}
Beam dump experiments, where weakly coupled particles are searched for in the collision of proton or electron beams with fixed targets, also provide limits on light pseudoscalars (often called axion-like particles in this context).  To be successfully probed, the particles under consideration need to decay visibly and live long enough that they escape the target, but not so long that they do not decay before passing the detectors downstream, typically tens of meters away from the target.  This results in mass-dependent exclusion bands in the pseudoscalar's lifetime.

Light pseudoscalars are produced in beam dump experiments either through the Primakoff process via their coupling to two photons or by being radiated via direct couplings to SM fermions as the beam constituents are stopped, in a manner analogous to bremsstrahlung.  Each mode typically provides an ${\cal O}(1)$ fraction of the total production cross section.  The final states that are considered in searches for light pseudoscalars include $\gamma\gamma$, $e^+e^-$, and $\mu^+\mu^-$.

In the model we consider, the pseudoscalars are predominantly produced by the Primakoff process since they do not directly couple to SM fermions and the dominant decay mode is $\gamma\gamma$.  In placing limits on our parameter space, however, we simply apply the experiments' reported exclusions~\cite{Bergsma:1985qz,*Riordan:1987aw,*Davier:1989wz,*Bross:1989mp} to generic axion-like particles, ignoring subtleties in the slightly different production cross sections and different branching fractions.  This approximation results in conservative exclusion regions since reducing the direct couplings to SM fermions should not make for more stringent limits.  In the region of parameter space of interest we find upper limits on the pseudoscalar lifetime, corresponding to prompt decays inside the targets, or, equivalently, upper limits on the inverse coupling $M$ in Eq.~(\ref{eq:L_int}).  The values of these upper limits tend to roughly coincide with those coming from the requirement that the pseudoscalar also decays promptly at the LHC.
\end{subsection}

%%%%%%%%%%%%%%%%%%%%%%%%%%%%%%%%%%%%%%%%%%%%%%%%%%%%%%%%%%%%%%%%%%%%%%%%%%%%%%%%

\begin{subsection}{Further Limits}
\label{sec:further}
The coupling of $a$ to SM fermions becomes negligible in the non-relativistic limit.  Therefore, the strict constraints light scalars and vectors are subject to from diffractive low energy neutron scattering on nuclei do not apply in this case.  For the same reason, these particles are also not constrained by measurements of D--P transitions in muonic Si and Mg.

Additionally, light pseudoscalars can also be produced in the decays of excited nuclear states in nuclear reactors.  However, this production mode is kinematically limited to $m_a\lsim 10\MeV$, and does not extend the exclusions we have considered.
\end{subsection}

%%%%%%%%%%%%%%%%%%%%%%%%%%%%%%%%%%%%%%%%%%%%%%%%%%%%%%%%%%%%%%%%%%%%%%%%%%%%%%%%

\begin{subsection}{Allowed Regions}
In Fig.~\ref{fig:constraints}, we show regions of $\left(M,m_a\right)$ excluded by the constraints outlined in Secs.~\ref{sec:g-2}--\ref{sec:further}.  We also display contours of $1~{\rm cm}$ and $50~{\rm cm}$ decay lengths, assuming $\gamma=m_h/2m_a$ with $m_h=125\GeV$.
\begin{figure}
\rotatebox{0}{\resizebox{80mm}{!}{\includegraphics{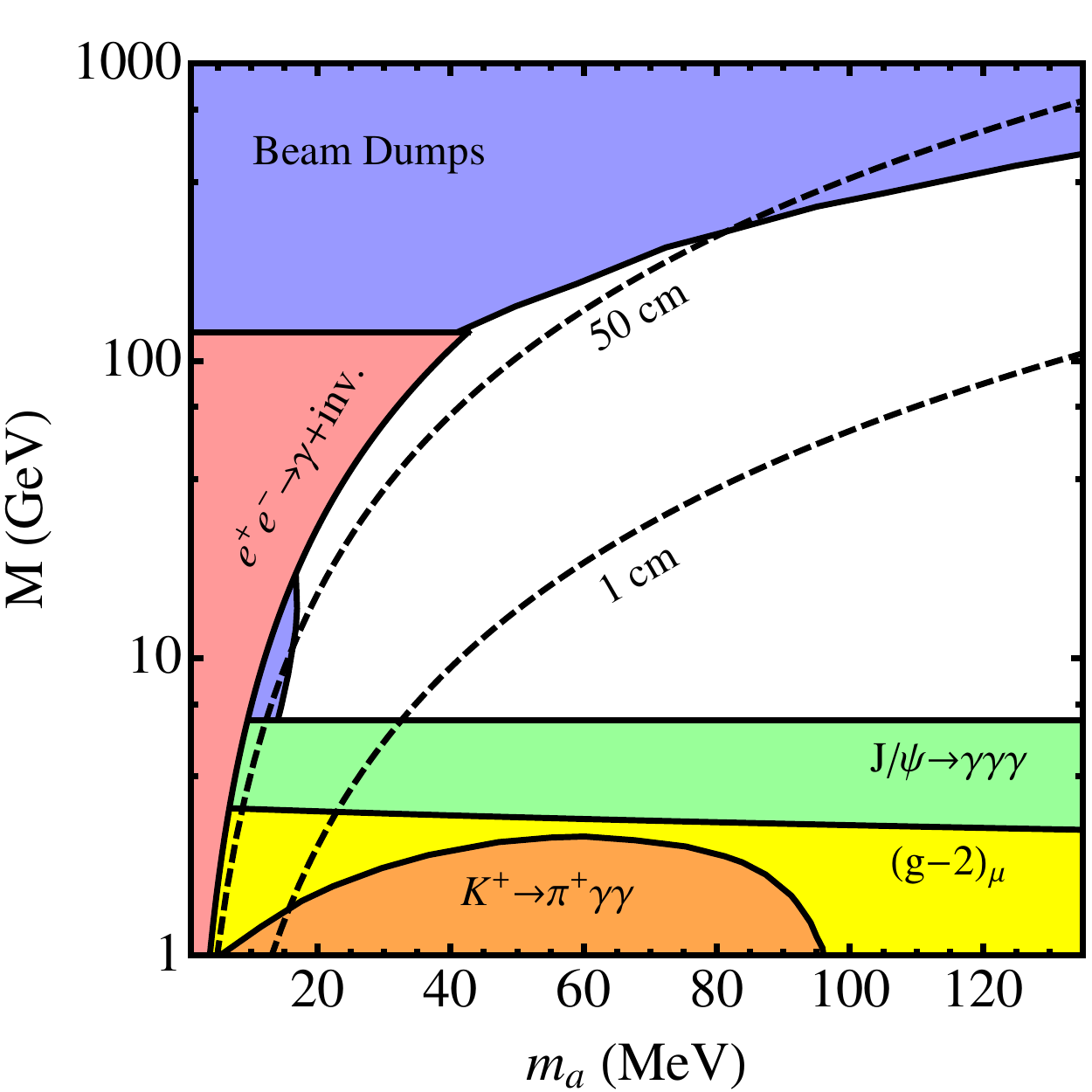}}}
\caption{Regions of pseudoscalar mass and inverse $a\gamma\gamma$ coupling $M$ excluded by the constraints in Secs.~\ref{sec:g-2}--\ref{sec:further}.  The dashed curves are contours of $1~{\rm cm}$ and $50~{\rm cm}$ decay lengths for the pseudoscalar, assuming a boost $\gamma=m_h/2m_a$ with $m_h=125\GeV$.}
\label{fig:constraints}
\end{figure}
\end{subsection}
\end{section}

%%%%%%%%%%%%%%%%%%%%%%%%%%%%%%%%%%%%%%%%%%%%%%%%%%%%%%%%%%%%%%%%%%%%%%%%%%%%%%%%
%%%%%%%%%%%%%%%%%%%%%%%%%%%%%%%%%%%%%%%%%%%%%%%%%%%%%%%%%%%%%%%%%%%%%%%%%%%%%%%%
%%%%%%%%%%%%%%%%%%%%%%%%%%%%%%%%%%%%%%%%%%%%%%%%%%%%%%%%%%%%%%%%%%%%%%%%%%%%%%%%

\begin{section}{Model Building}
\label{sec:models}
Having investigated the phenomenology associated with a light pseudoscalar coupled to the Higgs, we speculate on a few scenarios where the $aF\tilde F$ interaction in Eq.~(\ref{eq:L_int}) could result.  As we have seen, given a 125~GeV Higgs, $m_a$ needs to be on the order of tens of MeV for the photons from its decay to be collimated enough to plausibly fake a single photon at the LHC.  Using this fact in Eqs.~(\ref{eq:declength}) and (\ref{eq:scale_M}), we observe that $M$ cannot be at the TeV scale for $a$ to decay promptly as well to evade constraints from beam dump experiments.  Obtaining a  relatively low scale for a dimension-5 operator such as this involving interactions beyond the SM poses a model-building challenge.  

A simple scenario where an interaction $aF\tilde F$ is generated is if $a$ interacts with particles of mass $m$ that have electric charge $q$ in units of the electron's charge.  One can imagine that $a$ is either a composite made up of these charged particles or is fundamental and has a renormalizable coupling to them.  After integrating out these charged particles, the dimension-5 interaction between $a$ and two photons is generated with a scale $M\sim 4\pi^2 m/q^2$.  Given the requirement that $a$ decay promptly seems to require $a$ to couple to charged particles with a mass not much larger than tens of GeV.\footnote{We note that a scenario where $a\to\gamma\gamma$ is mediated by a coupling of $a$ to a new, light vector boson, $V$, that kinetically mixes with the photon (thereby lifting the requirement that $a$ couple to electrically charged particles) is not viable because the amplitude for such a process is proportional to $q_1^2 q_2^2\left(q_1^2-m_V^2\right)^{-1}\left(q_2^2-m_V^2\right)^{-1}$ where $q_{1,2}$ are the photons' momenta, which vanishes for on-shell photons. If $m_V$ vanishes, then $a\rightarrow VV$ decays will strongly dominate the branching ratios.}  Candidates for such charged particles are likely limited to SM fermions other than the $t$ quark.  Significant direct couplings to electrons or muons can be ruled out by anomalous magnetic moment measurements.  Couplings to heavy quarks $c$ and $b$ likely cause issues with well-studied quarkonium transitions.  We are left with $\tau$ and light quarks as SM mediators of the $aF\tilde F$ interaction.  For a recent analysis of new light bosons decaying to $\gamma\gamma$ see Ref.~\cite{McKeen:2011aa}.

%%%%%%%%%%%%%%%%%%%%%%%%%%%%%%%%%%%%%%%%%%%%%%%%%%%%%%%%%%%%%%%%%%%%%%%%%%%%%%%%

\begin{subsubsection}{Coupling to $\tau$}
\label{sec:tau}
We can also imagine that the effective $aF\tilde F$ interaction is due to a coupling of $a$ to $\tau$ leptons, writing this coupling as\footnote{We write the interaction in terms of a pseudoscalar instead of an axial vector coupling here for simplicity.}
\begin{align}
{\cal L}_{a\tau}=-ig_\tau a\bar \tau\gamma^5\tau.
\end{align}
This leads to an effective operator mediating $a\to\gamma\gamma$ in Eq.~(\ref{eq:L_int}) with (assuming $m_a\ll m_\tau$)
\begin{align}
M=\frac{4\pi^2m_\tau}{g_\tau}.
\end{align}

The experimental information about the anomalous magnetic moment of $\tau$~\cite{Abdallah:2003xd}
\begin{align}
-0.052<\frac{\left(g-2\right)_\tau}{2}<0.013,
\end{align}
limits $g_\tau\lsim 2.5$ for $m_a>10\MeV$.

A similar limit of $g_\tau\lsim 2-3$ also comes from the OPAL Collaboration's result~\cite{Acton:1991dq}
\begin{align}
{\cal B}\left(Z\to\tau^+\tau^-\gamma\right)<7.3\times10^{-4},
\end{align}
where $Z\to\tau^+\tau^-a\to\to\tau^+\tau^-\gamma\gamma$ can mimic $Z\to\tau^+\tau^-\gamma$ events~\cite{McKeen:2011aa}.

Without a dedicated search for these pseudoscalars in $\tau$ decays, it is difficult to be certain about constraints on this scenario coming from $\tau$ decay.  This is especially true if $a$ has a significant invisible width, a scenario which we discuss in the Appendix.  For example, the branching ratio of $\tau$ to $a$ and $\mu$ can be estimated to be~\cite{McKeen:2011aa}
\begin{align}
{\cal B}\left(\tau^\pm\to a \mu^\pm\bar\nu\nu\right)\simeq 3.5\times10^{-5}g_\tau^2.
\end{align}
Some of these decays could appear as $\tau^\pm\to\mu^\pm\bar\nu\nu\gamma$ decays, given the decay $a\to\gamma\gamma$.  Even with the large Yukawa allowed by $\left(g-2\right)_\tau$ and if all of the $\tau^\pm\to a \mu^\pm\bar\nu\nu$ were registered as $\tau^\pm\to\mu^\pm\bar\nu\nu\gamma$ decays, the contribution is still below the uncertainty on the measurement of this branching of $\left(3.61\pm0.38\right)\times10^{-3}$~\cite{Bergfeld:1999yh} which agrees well with expectations.  Limits from $\tau^\pm\to e^\pm\bar\nu\nu\gamma$ are less stringent and hadronic $\tau$ decays pose a bit more difficulty in prediction.

In this scenario, the scale $M$ is bounded by
\begin{align}
M=\frac{4\pi^2m_\tau}{g_\tau}\gsim 35\GeV.
\end{align}
\end{subsubsection}

%%%%%%%%%%%%%%%%%%%%%%%%%%%%%%%%%%%%%%%%%%%%%%%%%%%%%%%%%%%%%%%%%%%%%%%%%%%%%%%%

\begin{subsubsection}{Mixing with $\pi^0$}
\label{sec:pi0}
Coupling $a$ to light quarks so that $a$ mixes with $\pi^0$ could provide a solution.  Given the couplings (ignoring $s$ quarks for clarity)
\begin{align}
{\cal L}_{aq}=-ig_u a\bar u\gamma^5u-ig_d a\bar d\gamma^5d,
\end{align}
using the partially conserved axial current, one can estimate the mixing angle,
\begin{align}
\theta\sim\frac{\left(g_u-g_d\right)\left|\langle \bar q q\rangle\right|}{F_\pi\left(m_{\pi^0}^2-m_a^2\right)}\simeq\left(g_u-g_d\right)\frac{\left(400\MeV\right)^2}{\left(m_{\pi^0}^2-m_a^2\right)},
\end{align}
where $\langle \bar q q\rangle$ is the value of the light quark condensate and $F_\pi=92.2\MeV$ is the pion decay constant.  Here, the scale $M$ is given by its analogue in the $\pi^0$ case scaled by the mixing angle,
\begin{align}
M=\frac{4\pi^2F_\pi}{\sin\theta}.
\end{align}

If there are no symmetry conditions on $g_u$ and $g_d$, then we will in general expect additional isospin violation due to interactions involving $a$.  The relative size of these effects is of the order $g_{u,d}^2/\left(m_{u,d}^2/F_\pi^2\right)$.  If we demand that these not exceed $\sim 1\%$ then we require $g_{u,d}\lsim 10^{-3}$.  Furthermore, one must then contend with corrections to pion beta decay $\pi^\pm\to\pi^0e^\pm\nu$ which is shifted by
\begin{align}
\Gamma\left(\pi^\pm\to\pi^0e^\pm\nu\right)=\cos^2\theta~\Gamma_{\rm SM}\left(\pi^\pm\to\pi^0e^\pm\nu\right).
\end{align}
The $0.6\%$ agreement of theory and experiment for this rate~\cite{Pocanic:2003pf} implies that $\theta\lsim0.08$.  Such a bound is compatible with the limits above on the Yukawa couplings from isospin violation.

The current measurement of the $\pi^0$ lifetime is at the 3\% level~\cite{Larin:2010kq}, and therefore does not pose any stricter constraints on the strength of this mixing.  Furthermore, given a mixing angle of this size, the contribution to $\pi^0\to e^+e^-$ from $a$--$\pi^0$ mixing is negligible.

There are potentially strict limits in this scenario coming from the $K^+\to\pi^+\gamma\gamma$ limits described in Sec.~\ref{sec:mesons} since the rate for $K^+\to\pi^++a$ is enhanced.  The rate can be estimated as
\begin{align}
{\cal B}\left(K^+\to\pi^+a\right)&\simeq\sin^2\theta~{\cal B}\left(K^+\to\pi^+\pi^0\right)
\\
&=\sin^2\theta\times0.21.
\nonumber
\end{align}
The limits on this branching at the $10^{-7}$ level~\cite{Kitching:1997zj} imply that $\theta\lsim10^{-3}$, or equivalently, $M\gsim1\TeV$.  This does not allow for prompt decays at the LHC if $a$ is produced in the decay of a $125\GeV$ Higgs and is problematic with respect to beam dump constraints.  However, the analysis in Ref.~\cite{Kitching:1997zj} only selected events corresponding to $m_a<100\MeV$ because of the very large background from $K^+\to\pi^+\pi^0$.  Thus, to avoid the strict limit on $M$ that results from this analysis, $a$ must have a mass greater than $100\MeV$ if its coupling to photons is mediated by a mixing with $\pi^0$.

The limit from pion beta decay implies that $M\gsim 47\GeV$ in this case.
\end{subsubsection}

%%%%%%%%%%%%%%%%%%%%%%%%%%%%%%%%%%%%%%%%%%%%%%%%%%%%%%%%%%%%%%%%%%%%%%%%%%%%%%%%

\begin{subsubsection}{Other Scenarios}
\label{sec:othermodels}
In addition to the two scenarios described above we briefly mention a few other potential models that could lead to the interactions in Eq.~(\ref{eq:L_int}) with a scale $M$ on the order of tens to a few hundred GeV.  As in~\cite{Dobrescu:2000jt}, the pseudoscalar could be coupled to heavy vector-like matter that mediates its interaction with photons.  However, ensuring that the scale $M$ is small enough in this situation so that $a$ decays promptly is difficult.  This could be solved by giving the heavy matter a large electric charge or gauging it under a non-Abelian group with a large number of colors.  We also speculate that sterile neutrinos with very large transition magnetic moments coupled to $a$ could also offer a solution since their masses need not be at the weak scale.
\end{subsubsection}

%%%%%%%%%%%%%%%%%%%%%%%%%%%%%%%%%%%%%%%%%%%%%%%%%%%%%%%%%%%%%%%%%%%%%%%%%%%%%%%%

\end{section}

%%%%%%%%%%%%%%%%%%%%%%%%%%%%%%%%%%%%%%%%%%%%%%%%%%%%%%%%%%%%%%%%%%%%%%%%%%%%%%%%
%%%%%%%%%%%%%%%%%%%%%%%%%%%%%%%%%%%%%%%%%%%%%%%%%%%%%%%%%%%%%%%%%%%%%%%%%%%%%%%%
%%%%%%%%%%%%%%%%%%%%%%%%%%%%%%%%%%%%%%%%%%%%%%%%%%%%%%%%%%%%%%%%%%%%%%%%%%%%%%%%

\begin{section}{Outlook}
\label{sec:outlook}
As discussed in Sec.~\ref{sec:misid}, the scenario analyzed in this paper can be tested in $h\rightarrow\gamma\gamma$ events, since the expected rate depends sensitively on the photon identification criteria used in the analysis. Furthermore, the presence of more photons implies that the fraction of events with a photon conversion is higher than in a pure $\gamma\gamma$ sample. Although CMS has already presented the best-fit rates in conversion and unconverted events separately, the statistics remain low, and we leave such analysis for the future.  We also note that in the case where the $a$ is long-lived, $\gamma c\tau\sim50~{\rm cm}$, it is imaginable that the decay length may be determined from the distribution of conversion radii, even though since the states are highly boosted they do not display displaced vertices in the traditional sense.

The light pseudoscalar hypothesis can also be tested by looking for a $3\gamma$ signal with invariant mass matching that of the Higgs boson. If the Higgs is produced with a sizable boost, some decays will yield one slow $a$ and one fast $a$ in the lab frame, and the former will produce widely separated photons. Alternatively, if $m_a$ is much larger than the values studied here, the pseudoscalar can still contribute some events to $h\rightarrow 2\gamma$ if ${\cal B}(h\rightarrow aa)$ is large enough, but then the particle may be more easily seen in searches for $h\rightarrow 4\gamma$~\cite{Chang:2006bw}.

A different way to search for a light pseudoscalar coupling to photons is to produce the $a$ directly in Primakoff-type experiments, $\gamma+{\rm Nuc.}\to a+{\rm Nuc.}$, which would allow the mass of $a$ to be directly reconstructed.  A proposal for an upgrade to the PrimEx Experiment at Jefferson Lab to measure $\Gamma\left(\eta\to\gamma\gamma\right)$ envisions collisions of photons with an energy $E_\gamma\simeq11\GeV$ on a liquid hydrogen target with a luminosity of ${\cal L}\sim 10^{-2}~{\rm nb}^{-1}{\rm s}^{-1}$ with a run of 45 days~\cite{primex}.  Using the interaction in Eq.~(\ref{eq:L_int}), the cross section for the Primakoff production of $a$ in the collision of a photon with a proton can be found to be
\begin{align}
\frac{d\sigma}{dQ^2}\simeq\frac{2\pi^2\alpha^3}{M^2}\frac{\left|F_{\rm em}\left(Q^2\right)\right|^2}{Q^2},
\end{align}
where $Q$ is the momentum transferred to the proton and $F_{\rm em}$ is its electromagnetic form factor.  Using a simple dipole form factor and a proton charge radius of $0.87~{\rm fm}$ (the results do not depend sensitively on the form factor and charge radius---using $0.84~{\rm fm}$ as measured in muonic hydrogen does not change the estimate) and a detection efficiency of 60\%, this cross section would lead to
\begin{align}
N\left(a\right)\simeq 10^4\left(\frac{10\GeV}{M}\right)^2
\end{align}
pseudoscalars collected with $m_a=40\MeV$.  For comparison, there would be about $10^4$ $\eta$'s produced via the Primakoff process and subsequently decaying to two photons under the same conditions.

It appears likely, at least statistically, that pseudoscalars with parameters in the range that is interesting in the context of the $h\to\gamma\gamma$ signal can be probed at future Primakoff experiments.
\end{section}

%%%%%%%%%%%%%%%%%%%%%%%%%%%%%%%%%%%%%%%%%%%%%%%%%%%%%%%%%%%%%%%%%%%%%%%%%%%%%%%%
%%%%%%%%%%%%%%%%%%%%%%%%%%%%%%%%%%%%%%%%%%%%%%%%%%%%%%%%%%%%%%%%%%%%%%%%%%%%%%%%
%%%%%%%%%%%%%%%%%%%%%%%%%%%%%%%%%%%%%%%%%%%%%%%%%%%%%%%%%%%%%%%%%%%%%%%%%%%%%%%%

\begin{section}{Conclusions}
\label{sec:conclusions}
In this paper we have investigated a simple model that can give rise to an apparent excess relative to the SM in the $h\to\gamma\gamma$ channel.  In our study the two photon excess comes from Higgs decays to two light, boosted pseudoscalars, $h\to aa$, and each pseudoscalar decays into two photons.  For very light pseudoscalars, each pair of photons is highly collimated and a non-negligible fraction of the pairs can appear as single photons, even at high-resolution detectors like those at the LHC.  This scenario serves as an example where photon jets~\cite{Toro:2012sv} are produced and helps to motivate the future experimental study of these objects in more detail. 

We have estimated the fraction of $4\gamma$ events that appear as $2\gamma$, taking into account the fine granularity of the ATLAS detector and the tight photon selection criteria.  Additionally, we have investigated subtleties that may arise when one photon or more converts.  We have assumed that the different analyses used at CMS and ATLAS do not give rise to large deviations between the $h\to\gamma\gamma$ measurements, but we note that the amount of $4\gamma$ contamination can be analysis-dependent and thus the differing analyses may be used as a test of the hypothesis.

In addition to a diphoton excess, the model also predicts a deficit relative to the SM for Higgs decays into other final states.  Although the uncertainties are large, the data that have been collected so far at the LHC and Tevatron are consistent with both of these predictions, and with more data the model will be easily tested. The relative size of a potential excess in the $\gamma\gamma$ channel and a decrement in the remaining channels will be a direct probe of the mass of the pseudoscalar, with a lighter pseudoscalar better able to accommodate a larger signal in the non-$\gamma\gamma$ channels for a fixed $\gamma\gamma$ rate.  Lower bounds around $m_a\simeq10\MeV$ from $\gamma+{\rm inv.}$ searches and beam dump experiments offer a complimentary sensitivity to the model.

Finally, the mass of the light pseudoscalar can be well-measured at Primakoff experiments. As we discussed in Sec.~\ref{sec:outlook}, future Primakoff experiments appear well-poised to test models involving new light bosons coupled to photons like the one we have considered, underscoring the complementarity of experiments at the intensity frontier with high-energy studies of the Higgs boson.   
\end{section}

%%%%%%%%%%%%%%%%%%%%%%%%%%%%%%%%%%%%%%%%%%%%%%%%%%%%%%%%%%%%%%%%%%%%%%%%%%%%%%%%
%%%%%%%%%%%%%%%%%%%%%%%%%%%%%%%%%%%%%%%%%%%%%%%%%%%%%%%%%%%%%%%%%%%%%%%%%%%%%%%%
%%%%%%%%%%%%%%%%%%%%%%%%%%%%%%%%%%%%%%%%%%%%%%%%%%%%%%%%%%%%%%%%%%%%%%%%%%%%%%%%

\begin{acknowledgments}
The authors would like to thank J.~Albert, V.~Bansal, K.~Jensen, M.~Lefebvre, J.~Mitrevski, J.~Nielsen, M.~Pospelov, J.~Redondo, and A.~Ritz for helpful discussions.  The work of DM was supported by NSERC, Canada.  PD is supported by the DOE under Grant No. DE-FG02-04ER41286.
\end{acknowledgments}

%%%%%%%%%%%%%%%%%%%%%%%%%%%%%%%%%%%%%%%%%%%%%%%%%%%%%%%%%%%%%%%%%%%%%%%%%%%%%%%%
%%%%%%%%%%%%%%%%%%%%%%%%%%%%%%%%%%%%%%%%%%%%%%%%%%%%%%%%%%%%%%%%%%%%%%%%%%%%%%%%
%%%%%%%%%%%%%%%%%%%%%%%%%%%%%%%%%%%%%%%%%%%%%%%%%%%%%%%%%%%%%%%%%%%%%%%%%%%%%%%%
\appendix*
\section{Invisible Decays}
\label{sec:appendix}
In this Appendix, we describe a simple extension of the model in Sec.~\ref{sec:model} to allow for $a$ to decay invisibly.  The Higgs phenomenology and direct constraints change somewhat from the case where $a$ decays purely to $\gamma\gamma$.  As before, we present fits to the Higgs data and show allowed regions of parameter space.

%%%%%%%%%%%%%%%%%%%%%%%%%%%%%%%%%%%%%%%%%%%%%%%%%%%%%%%%%%%%%%%%%%%%%%%%%%%%%%%%

\begin{subsection}{Framework}
In addition to the two photon decay channel, the pseudoscalar $a$ could have an appreciable invisible decay width.  As a concrete example, we can add an interaction with a stable (on collider scales) SM singlet Dirac fermion $\chi$,
\begin{align}
{\cal L}_{a\chi}&=\frac{\partial_\mu a}{M^\prime}\,\bar\chi \gamma^\mu\gamma^5\chi,
\end{align}
where $M^\prime$ is another scale describing the strength of this interaction.  The rate for $a\to\chi\bar\chi$ is then
\begin{align}
\Gamma\left(a\to\bar\chi\chi\right)=\frac{m_\chi^2m_a}{8\pi M^{\prime2}}\beta_\chi
\end{align}
When this channel is open, the branching ratio for $a\to\gamma\gamma$ becomes
\begin{align}
{\cal B}\left(a\to\gamma\gamma\right)&=\left[1+\frac{\beta_\chi}{2}\left(\frac{1}{\pi\alpha}\frac{m_\chi M}{m_a M^\prime}\right)^2\right]^{-1},
\end{align}
where $\beta_\chi^2=1-4m_\chi^2/m_a^2$.  The decay length (for a boost $\gamma=m_h/2m_a$) is shortened by a factor of this branching ratio,
\begin{align}
\gamma c\tau&\simeq=1.15~{\rm mm}~\left(\frac{M}{10\GeV}\right)^2\left(\frac{m_a}{40\MeV}\right)^{-4}
\\
&~~~~~~~~~~~~~~~~~~~~~~~~\times\left(\frac{m_h}{125~{\rm GeV}}\right)\left(\frac{{\cal B}\left(a\to\gamma\gamma\right)}{0.1}\right),
\nonumber
\end{align}
and the scale $M$ can then be expressed as
\begin{align}
M&=29~{\rm GeV}~\left(\frac{\gamma c\tau}{1~{\rm cm}}\right)^{1/2}\left(\frac{m_a}{40\MeV}\right)^{2}
\label{eq:scale_M_inv}
\\
&~~~~~~~~~~~~~~~~\times\left(\frac{m_h}{125\GeV}\right)^{-1/2}\left(\frac{{\cal B}\left(a\to\gamma\gamma\right)}{0.1}\right)^{-1/2}.
\nonumber
\end{align}
For a fixed decay length, a sizable invisible branching fraction for $a$ allows the scale $M$ to be somewhat larger than it would be in the case where $a$ only decays to photons.
\end{subsection}

%%%%%%%%%%%%%%%%%%%%%%%%%%%%%%%%%%%%%%%%%%%%%%%%%%%%%%%%%%%%%%%%%%%%%%%%%%%%%%%%

\begin{subsection}{Higgs Signal}
The primary changes to Higgs phenomenology are in ${\cal B}\left(h\to\gamma\gamma\right)_{\rm eff}$ and $R_{\gamma\gamma}$, now given by
\begin{align}
{\cal B}\left(h\to\gamma\gamma\right)_{\rm eff}&={\cal B}\left(h\to\gamma\gamma\right)
\\
&~~~~+\epsilon{\cal B}\left(h\to aa\right){\cal B}\left(a\to\gamma\gamma\right)^2
\nonumber
\end{align}
and 
\begin{align}
R_{\gamma\gamma}=1+{\cal B}\left(h\to aa\right)\left(\frac{\epsilon{\cal B}\left(a\to \gamma\gamma\right)^2}{{{\cal B}_{\rm SM}\left(h\to\gamma\gamma\right)}}-1\right).
\end{align}

\begin{figure}
\rotatebox{0}{\resizebox{80mm}{!}{\includegraphics{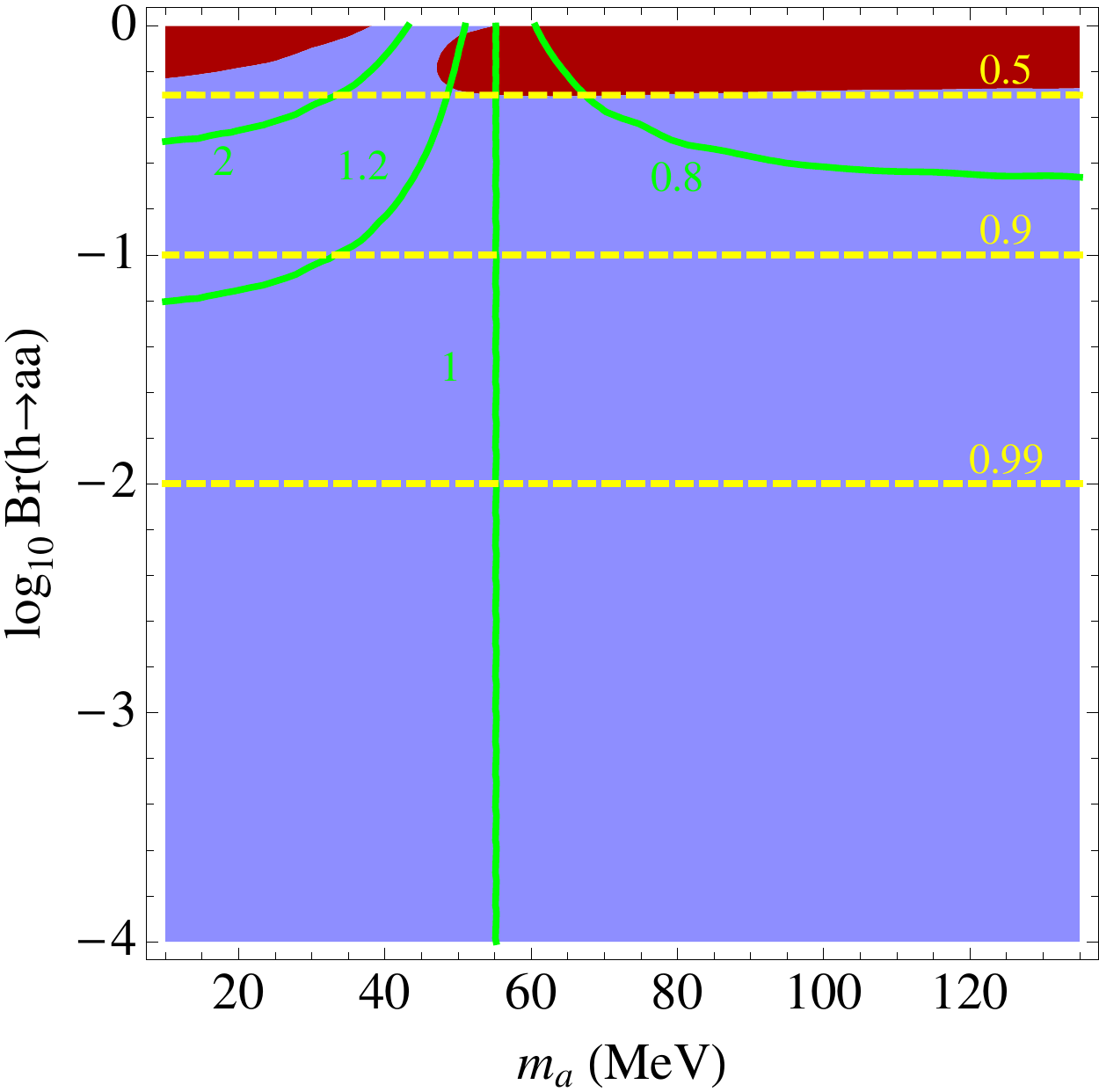}}}
\caption{Regions of light pseudoscalar parameter space that are favored (blue/light) and disfavored (red/dark) by the current best-fit signal strengths in the $h\rightarrow\gamma\gamma,ZZ,WW,bb,\tau\tau$ channels. Contours are overlaid for the net diphoton (solid green) and $ZZ,WW,bb,\tau\tau$ rates (dashed yellow) expected at the LHC relative to the SM rates. ${\cal B}\left(a\to \gamma\gamma\right)$ is fixed to $0.1$, indicating a 90\% branching of $a\rightarrow \rm{invisible}$.}
\label{fig:exclappx}
\end{figure}

Fixing ${\cal B}\left(a\to \gamma\gamma\right)=0.1$, we regenerate the exclusion contours and contours of fixed $R_{\gamma\gamma,XX}$ in Fig.~\ref{fig:exclappx}. The best-fit $\chi^2/{\rm d.o.f.}=6.5/10$, and the best-fit point is close to the intersection of the $R_{\gamma\gamma}=2$ and $R_{XX}=0.5$ contours. Like the ${\cal B}\left(a\to \gamma\gamma\right)=1$ case, the $\chi^2$ is shallow and the best-fit point is not yet a meaningful quantity, so we do not show it in Fig.~\ref{fig:exclappx}.

In addition to an increased apparent branching to $\gamma\gamma$, there are contributions to the invisible width of the Higgs, which can be related to the increased diphoton branching ratio,
\begin{align}
{\cal B}\left(h\to{\rm invisible}\right)&={\cal B}\left(h\to aa\right){\cal B}\left(a\to\bar\chi\chi\right)^2
\\
&={\cal B}\left(h\to aa\right)\left[1-{\cal B}\left(a\to\gamma\gamma\right)\right]^2.
\nonumber
\end{align}
There is also a ``monophoton" branching,
\begin{align}
&{\cal B}\left(h\to\gamma+{\rm invisible}\right)={\cal B}\left(h\to aa\right){\cal B}\left(a\to\gamma\gamma\right)
\\
&~~~~~~~~~~~~~~~~~~~~~~~~~~~~~~~~\times{\cal B}\left(a\to\bar\chi\chi\right)\times2\epsilon'
\nonumber
\\
&=2\epsilon'\,{\cal B}\left(h\to aa\right){\cal B}\left(a\to\gamma\gamma\right)\left[1-{\cal B}\left(a\to\gamma\gamma\right)\right],
\end{align}
where $\epsilon'$ is the efficiency for the one photon jet in the event to be reconstructed as a single photon.
Future LHC monophoton searches with low MET cuts may be able to probe this channel.
\end{subsection}

%%%%%%%%%%%%%%%%%%%%%%%%%%%%%%%%%%%%%%%%%%%%%%%%%%%%%%%%%%%%%%%%%%%%%%%%%%%%%%%%

\begin{subsection}{Direct Constraints on $a$}
We now analyze changes to the direct limits given in Secs.~\ref{sec:g-2}--\ref{sec:further} that arise if $a$ has an appreciable invisible branching fraction.

\begin{figure}
\rotatebox{0}{\resizebox{80mm}{!}{\includegraphics{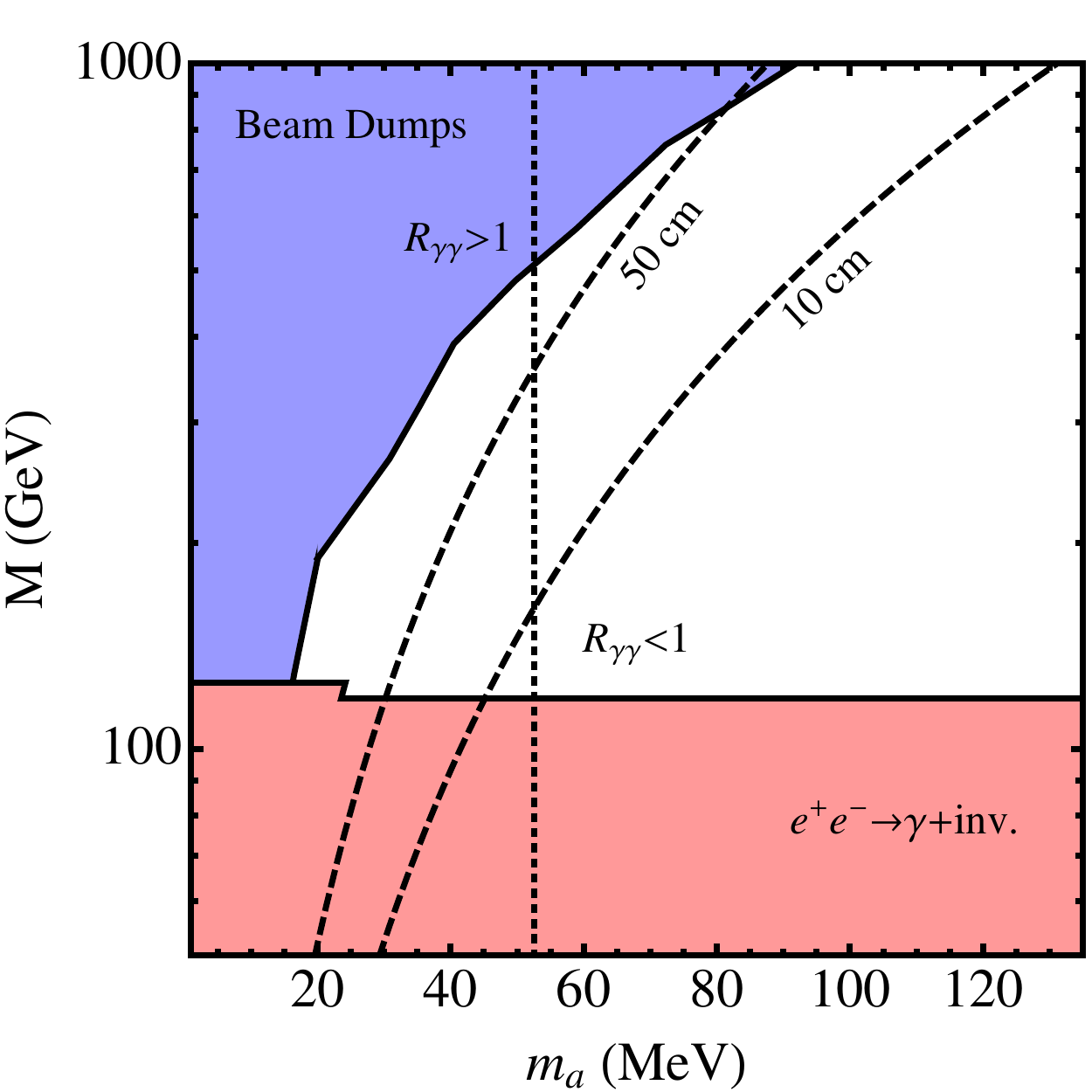}}}
\caption{Same as Fig.~\ref{fig:constraints} in the case that ${\cal B}\left(a\to\gamma\gamma\right)=0.1$.  The dashed curves are contours of 10 and 50~cm decay lengths for $\gamma=m_h/2m_a$ and $m_h=125\GeV$.  To the left of the vertical dotted line, $R_{\gamma\gamma}>1$ [Eqs.~(\ref{eq:Rgaga}) and (\ref{Rdef})] while to its right, $R_{\gamma\gamma}<1$.  In this case, only the beam dump and $e^+e^-\to\gamma+{\rm inv.}$ limits, which are the most constraining, are shown.}
\label{fig:constraints_inv}
\end{figure}
Signals involving missing energy now apply regardless of $a$'s decay length.  The $e^+e^-\to\gamma+{\rm inv.}$ limit then becomes the strongest lower limit on $M$.

The limits on $M$ from beam dumps must be rescaled by a factor of ${\cal B}\left(a\to\gamma\gamma\right)^{-1/2}$ since they are limits on the $a$ lifetime.  We conservatively ignore any shrinking of the exclusion region due to the smaller visible branching ratio.

An invisibly decaying $a$ could contribute to star cooling by providing an additional channel for energy to leave a star.  However, such limits are not important in the mass range that we consider here.

If $a$ mixes with the $\pi^0$ and can decay invisibly, there are stringent limits on the invisible decay rate of the $\pi^0$, with the collider upper limit measured to be $2.7\times10^{-7}$~\cite{Artamonov:2005cu} and a limit from Big Bang Nucleosynthesis that is several orders of magnitude stronger~\cite{Lam:1991bm}.  These are quite constraining and make the scenario where the $aF\tilde F$ coupling is generated by a mixing with $\pi^0$ described in Sec.~\ref{sec:pi0} unlikely if $a$ has an appreciable invisible branching fraction.

A large invisible branching fraction for $a$ renders the model where the coupling to photons is mediated by a coupling to $\tau$ more plausible. This is because $\tau$ decays necessarily involve at least one neutrino in the final state so observing missing energy in a $\tau$ decay does not signal new physics or a rare SM decay.

We reproduce the low-energy exclusions of Fig.~\ref{fig:constraints} in Fig.~\ref{fig:constraints_inv} for ${\cal B}\left(a\to\gamma\gamma\right)=0.1$, showing only the most stringent limits from beam dumps and $e^+e^-\to\gamma+{\rm inv}$.
\end{subsection}

\bibliography{h_to_4A}
\end{document}